\documentclass[11pt]{iopart}

\newcommand{\xxi}[1]{\stackrel[#1]{}{\xi}}
\usepackage{iopams} 
\usepackage{stackrel}
\newcommand{\be}{\begin{equation}}
\newcommand{\ee}{\end{equation}}
\newcommand{\beal}{\begin{align}}
\newcommand{\eeal}{\end{align}}
\newcommand{\nba}{\numparts \begin{eqnarray}} 
\newcommand{\nea}{\end{eqnarray} \endnumparts}
\newcommand{\ba}{\begin{eqnarray}}
\newcommand{\ea}{\end{eqnarray}}

\newcommand{\bea}{\begin{eqnarray}}
\newcommand{\eea}{\end{eqnarray}}

\newcommand{\nn}{\nonumber}
\eqnobysec

\newcommand{\ben}{\begin{equation*}}
\newcommand{\een}{\end{equation*}}
\newcommand{\bean}{\begin{eqnarray*}}
\newcommand{\eean}{\end{eqnarray*}}

\begin{document}

\title[]{Null hypersurfaces in general relativity: Intrinsic symmetries and differential invariants}

\author{G. Dautcourt}

\address{}
\ead{dautcourt@germanynet.de}
\vspace{10pt}
\begin{indented}
\item[]             
\end{indented}

\begin{abstract}
This paper investigates intrinsic Killing symmetries of null hypersurfaces $\mathcal{N}_3$ within the framework of general relativity.
To this end we consider $\mathcal{N}_3$ as detached from the embedding spacetime 
and equipped with a degenerate metric of signature (0,+,+). As geometrical 
tools we use a triad calculus and differential invariants. 
Based on prior work, we  present a classification of null hypersurfaces 
according to groups of motion up to the fourth order. For each type certain normal forms of the metric are given, and their invariants are listed. A discussion of horizons - defined as null hypersurfaces with vanishing shear and divergence - is included. 
\end{abstract}

\section{Introduction}\label{sec1}

Null hypersurfaces, also referred to as lightlike, characteristic or isotropic hypersurfaces, play a central role in general relativity. Prominent examples include the formation and structure of horizons in black hole physics, the theory of gravitational radiation with characteristic propagation fronts, and the cosmological past light cone, which encodes the accessible empirical information of the Universe.                        
Another significant example is the characteristic initial value problem of 
Einstein's field equations, based on a pair of null hypersurfaces, which was first addressed many years ago \cite{pen0,newman,daut1}. Since then, many further 
approaches for studying null hypersurfaces and their embedding in spacetime have been developed, see the extensive reference lists in recent publications \cite{chandra,odak,ashtekar,manz2,mars1,blitz,alvarez,manz1}. 

Null hypersurfaces are embedded within the spacetime geometry, and geometrical quantities associated with this embedding - such as surface gravity and rigged null directions - play a crucial role in various applications. However, a clear geometric understanding can only be achieved by carefully separating intrinsic properties from embedding characteristics. Thus the null hypersurface can initially be considered as an independent object, detached from the ambient spacetime (see also \cite{mars1}). 

We adopt this strategy and investigate in this paper the \emph{Killing isometries of a three-dimensional standalone null hypersurface $\mathcal{N}_3$}. A related study was carried out previously \cite{daut4}, but it contained many misprints and errors. Here we correct that analysis.

The article is organized as follows.
In section 2 an elementary presentation of the geometry of null surfaces is given in terms of a \emph{triad formalism}. A detailed derivation of differential invariants follows in section 3. Section 4 treats the Killing equation and its integrability conditions in the triad calculus.
Section 5 focused on hypersurfaces admitting a null Killing vector (Killing horizons).
Section 6 determines metrics admitting groups of motion $G_1$ and $G_2$.  
Sections 7 and 8 treat the same problem for the groups $G_3$ and $G_4$ respectively. In section 9 some properties of the metrics are briefly outlined, such as orbits of Killing vectors.                          
Appendix A summarizes the classification of null hypersurfaces in terms of 
differential invariants and provides tools for most of the calculations in the article.  Appendix B contains an overview of the found metrics.    

The considerations in this article refer to the 
\emph{inner} geometry, which is based solely on a degenerate metric $\gamma_{ik}$.
In \cite{pen1} and \cite{daut2,daut3} it was shown, that a whole hierarchy of null surface geometries exist. They range from the conformal geometry to specialized inner geometries, if less or more properties of the ambient spacetime are added. 
Another paper may address these issues.

More mathematically oriented treatments of degenerate metrics with isometries 
can be found in numerous publications, for example 
\cite{bekkara1},\cite{bekkara2},\cite{palomo}.

\section{A triad formalism for $\mathcal{N}_3$}

 We use a triad calculus to study null spaces $\mathcal{N}_3$.
$\mathcal{N}_3$ is defined as three-dimensional manifold with a smooth symmetric covariant tensor field $\gamma_{ik}(x_i)$ of matrix rank 2 and 
 signature $(0,1,1)$ as metric. Regular points $x_i$ 
 are defined by $rank~ \gamma_{ik}(x_i) =2.$ The rank may drop to 1 at focal points and 0 at a vertex, but such cases will not be considered. Also global aspects of $\mathcal{N}_3$ are not discussed here. We mention only that the special role of null directions suggests a product topology, e.g. $R^2\times R$ or $S^2\times R$.
 Since $\gamma_{ik}$ is degenerate, the equation
 \be
\gamma_{ik}\epsilon^k = 0
 \ee
 has a non-vanishing solution $\epsilon^k$, defined up to a factor. $\epsilon^k$ is called null direction or generator direction. The tangential lines of $\epsilon^k$ provide a foliation of $\mathcal{N}_3$.
 In addition to $\epsilon^k$ there are two further vectors $p^k,~q^k$,
 such that 
 \be
 p^k\gamma_{kl} = p_l \neq 0, ~ q^k\gamma_{kl}=q_l \neq 0.
 \ee
We combine $p^k$ and $q^k$ into a complex pseudo null vector:
\be
t^k = \frac{1}{\sqrt{2}}(p^k + iq^k)
\ee
The corresponding complex-conjugate vector is denoted by $\bar{t^k} = 
\frac{1}{\sqrt{2}}(p^k - iq^k)$. Together with the contravariant triad $\epsilon^k,t^k,\bar{t^k}$ we introduce a corresponding covariant triad $\gamma_k,t_k,\bar{t_k}$ by
\be
t^kt_k=0,~t^k\bar{t_k}=1,~\epsilon^kt_k=t^k\gamma_k=0,~ \epsilon^k\gamma_k =1.
\ee
The metric can now be written
\be
 \gamma_{ik} = t_i\bar{t_k} + \bar{t_i}t_k. 
\ee
Since $\gamma_{ik}$ is degenerate, there is no reciprocal contravariant metric $\epsilon^{ik}$ with $\epsilon^{ik}\gamma_{kl}=\delta^i_l$. For a non-degenerate metric the latter condition is equivalent to 
\bea
\gamma_{li}\gamma_{mk}\epsilon^{ik} = \gamma_{lm}. \nonumber
\eea
In the case of degeneration, this equation remains meaningful, but the solution for $\epsilon^{ik}$ is not unique, since 
\bea\nonumber
\epsilon'^{ik}  = \epsilon^{ik} + a^i\epsilon^k+a^k\epsilon^i
\eea
is also a solution for every $a^i$. Nevertheless, with 
\be
\epsilon^{ik} = \bar{t^i}t^k + \bar{t^k}t^i
\ee
 a contravariant metric $\epsilon^{ik}$ can be unambiguously assigned to each triad. $\epsilon^{ik}$ has the same rank as $\gamma_{ik}$ and is related to the covariant metric by
\be
\epsilon^{kl}\gamma_{il} = \delta^k_i - \epsilon^k \gamma_i.
\ee

The triad is not unique. It can be transformed into
\numparts
\bea \label{trga}
t'^i = e^{i\omega}(t^i - \lambda\bar{\kappa}\epsilon^i), \\ \label{trgb}
\epsilon'^i = \lambda\epsilon^i, \\ \label{trgc}
t'_i = e^{i\omega}t_i, \\ \label{trgd}
\gamma'_i= \frac{1}{\lambda}\gamma_i+ \kappa t_i + \bar{\kappa}\bar{t}_i, 
\eea
\endnumparts
$\omega, \lambda$ real, $\kappa$ complex. This transformation group 
is locally isomorphic to a 4-parameter subgroup of the Lorentz transformations that leave the null direction invariant. It splits into null rotations ($\kappa$), spacelike rotations ($\omega$) and a scale change ($\lambda$).

The introduction of covariant derivatives requires the definition of an affinity. This is well known to encounter difficulties in degenerate spaces due to the non-existence of a contravariant metric \cite{lemmer,katsuno}. We solve this
problem by introducing a \emph{class} of affinities, that is a triad-depending 
affinity (see \cite{daut2},\cite{daut3}). For a given triad we define the affinity by 
\be \label{aff}
\Gamma^l_{ik} = \frac{1}{2}\epsilon^{lm}(\gamma_{mi,k}+\gamma_{mk,i}-\gamma_{ik,m}) + \frac{1}{2}\epsilon^l(\gamma_{i,k}+\gamma_{k,i}).
\ee
Under general coordinate transformation
\be \label{tr0}
x'_i =x'_i(x_k)
\ee
with $det|\frac{\partial x'_i}{\partial x_k}| \neq 0$,
$\Gamma^i_{kl}$  behaves as affinity indeed.   
Using this affinity, the covariant derivatives of the triad (denoted by a 
stroke, for the ordinary derivative we use a comma) may be represented as:
\bea
\epsilon^i_{|k} = -(\rho \bar{t}_k + \sigma t_k)t^i 
-(\rho t_k+ \bar{\sigma}\bar{t}_k)\bar{t}^i +\chi t_k\epsilon^i+\bar{\chi}\bar{t}_k\epsilon^i, \\
t^i_{|k} = (\tau t_k -\bar{\tau}\bar{t}_k +i\nu\gamma_k)t^i
 +(i\varphi t_k-\bar{\chi}\gamma_k)\epsilon^i, \\
\gamma_{i|k} = i\varphi(t_i\bar{t}_k -\bar{t}_it_k) 
+ \chi(t_i\gamma_k-\gamma_it_k)
+ \bar{\chi}(\bar{t}_i\gamma_k-\gamma_i\bar{t}_k)
\\
t_{i|k} = \tau t_it_k - \bar{\tau}t_i\bar{t}_k +i\nu\gamma_kt_i 
+ \rho\gamma_it_k + \bar{\sigma}\bar{t}_k \gamma_i.
\eea
The so-called rotation coefficients $\rho~,\sigma~,\nu~,\tau~,\chi~,\varphi$
can also be displayed without affinity:
\bea
\rho+ i\nu = \epsilon^it^k(\bar{t}_{i|k}-\bar{t}_{k|i})
= \epsilon^it^k(\bar{t}_{i,k}-\bar{t}_{k,i}), \\ 
\sigma = \epsilon^i\bar{t}^k\bar{t}_{i|k} 
= \epsilon^i\bar{t}^k(\bar{t}_{i,k}- \bar{t}_{k,i}), \\
\tau = \bar{t}^i\bar{t}^kt_{i|k} =
\bar{t}^it^k(\bar{t}_{i,k} -\bar{t}_{k,i}), \\
\chi = \bar{t}^i\epsilon^k\gamma_{i|k} = \frac{1}{2}\bar{t}^i\epsilon^k( \gamma_{i,k}-\gamma_{k,i}), \\
i\varphi = \bar{t}^it^k\gamma_{i|k} = \frac{1}{2}\bar{t}^it^k(\gamma_{i,k}-\gamma_{k,i}).
\eea
It is appropriate to introduce directional differential operators:
\be
\delta =t^i\partial_i,~ \bar{\delta}= \bar{t}^i\partial_i,~ D=\epsilon^i\partial_i,~ \partial_i \equiv \frac{\partial}{\partial x_i}.
\ee
They satisfy the commutator relations
\bea \label{commu1}
D\delta -\delta D = (\rho+i\nu)\delta +\bar{\sigma}\bar{\delta} -2\bar{\chi}D, \\
D\bar{\delta} -\bar{\delta} D = (\rho-i\nu)\bar{\delta} +\sigma \delta 
-2\chi D, \\ \label{commu2}
\delta \bar{\delta} - \bar{\delta}\delta = 
\bar{\tau}\bar{\delta} - \tau\delta  -2i\varphi D. \label{commu3}
\eea
We also note the identities
\bea  \label{iiden}
D\tau &=& \bar{\delta}(\rho +i\upsilon)-\delta\sigma +\tau(\rho-i\upsilon)
+\sigma\bar{\tau}-2\chi(\rho+i\upsilon)+2\sigma\bar{\chi}, \\
iD\varphi &=&
\delta\chi-\bar{\delta}\bar{\chi}-\chi\bar{\tau}+\bar{\chi}\tau+2i\varphi\rho.
\eea
Under the triad transformations \eref{trga}-\eref{trgd}  the rotation coefficients transform as
\bea
\rho' &=& \lambda\rho, \\
\sigma' &=& \lambda e^{-2i\omega}\sigma, \\
\tau' &=& e^{-i\omega}(\tau +i\bar{\delta}\omega -i\kappa\lambda D\omega
-i\kappa\lambda\upsilon +\bar{\kappa}\lambda\sigma -\kappa\lambda\rho), \\
\upsilon' &=& \lambda\upsilon +\lambda D\omega, \\
\chi' &=& \frac{1}{2}e^{-i\omega}(2\chi+ \frac{\bar{\delta}\lambda}{\lambda}
-\kappa\rho\lambda +i\upsilon\kappa\lambda
-\bar{\kappa}\lambda\sigma +\lambda D\kappa),\\
i\varphi' &=& \frac{1}{\lambda}i\varphi +\frac{1}{2}(\bar{\kappa}\tau
-\kappa\bar{\tau}+\delta\kappa -\bar{\delta}\bar{\kappa})
 +\frac{1}{2}\kappa(2\bar{\chi}+\frac{\delta\lambda}{\lambda}
 -i\bar{\kappa}\lambda\upsilon -\kappa\lambda\bar{\sigma}
+\lambda D\bar{\kappa}) \nn \\
&& -\frac{1}{2}\bar{\kappa}(2\chi+\frac{\bar{\delta}\lambda}{\lambda}
 +i\kappa\lambda\upsilon -\bar{\kappa}\lambda\sigma+\lambda D\kappa)
\eea

We may use the transformations to simplify many calculations. Thus $\gamma_i$ can be chosen as gradient:
\be \label{simp0}
\gamma_i =  v_{,i.}.
\ee
In this case 
\be \label{simp}
\chi = \varphi = 0.
\ee
$\mathcal{N}_3$ is then covered by a set of spacelike 2-slices $v=const$.
The slices have a two-dimensional metric with a Gaussian curvature
\be \label{kkk} 
K= -2\tau\bar{\tau} + \bar{\delta}\bar{\tau} +\delta\tau.
\ee
The condition \eref{simp} is preserved under triad transformations with coefficients $\lambda$ and $\kappa$ restricted by
\bea
\delta\lambda = \lambda^2(-D\bar{\kappa} +i\bar{\kappa}\nu +\bar{\kappa}\rho +\kappa\bar{\sigma}), \\
\bar{\delta}\bar{\kappa} - \delta\kappa = \bar{\kappa}\tau - \kappa\bar{\tau}.
\eea
In general, a triad transformation leads to a \emph{different} set of space-like slices with a \emph{changed} Gaussian curvature, thus \emph{$K$ is in general no differential invariant of $\mathcal{N}_3$} \cite{daut2,daut3}. An important exception are horizons, cf. Section 4.

Another quantity which can be made to disappear in regular points is $\upsilon=\
\frac{1}{i}t_{i\mid k}\epsilon^k\bar{t}^i$,  which gives the deviation of the 
displacement of the components $t^i$ in the null direction from parallel transport with respect to the affinity $\Gamma^l_{~ik}$. 
$\upsilon=0$  is equivalent to parallel displacement of both the covariant and the contravariant triad along $\epsilon^i$. To  maintain $\upsilon=0$, the triad transformations would have to be restricted by the further requirement
\be
D\omega =0.
\ee 
In this paper we use mainly the simplified form of the triad formalism with $\chi= \varphi =\upsilon =0.$

It is also useful to introduce \emph{adapted coordinates} defined by $\epsilon^i=\delta^i_1$, using the transformations \eref{tr0}\footnote{One example where \emph{all} components of $\gamma_{ik}$ are required, is the null quasispherical gauge by Bartnik \cite{bartnik}.} 
. Then, from (2.1), $\gamma_{i1}=0$, so that the metric can be written in the standard form 
\be \label{gl1}
ds^2 = E(x_1,x_2,x_3)dx_2^2 +2 F(x_1,x_2,x_3)dx_2dx_3 + G(x_1,x_2,x_3)dx_3^2
\ee
with three functions $E, F, G.$ Adapted coordinates remain adapted if the transformations \eref{tr0} are restricted to (capital letters always run from 2 to 3)
\be \label{tr1a}
x'_1 = x'_1(x_1,x_A),
\ee
\be  \label{tr1b}
x'_A = x'_A(x_B).
\ee

\section{Differential invariants}               

Differential invariants are an important geometric element of null spaces. The invariants are functions of the metric and its derivatives up to a certain order:
\be
f = f(\gamma_{ik},~\gamma_{ik,l},~\gamma_{ik,lm}, ~ ...),
\ee
that are invariant under the transformation \eref{tr0}.
We can also represent the invariants by rotation coefficients and their 
directional derivatives. In this paper we consider only invariants up to 
second order and write therefore 
\be
f = f(\rho,\sigma,\tau,\nu,\chi,\varphi,D\rho, ...\bar{\delta}\varphi).
\ee
The function $f$ must be invariant under the triad transformations \eref{trga}-\eref{trgd}.
We use the restricted set of rotation coefficients and will first consider 
first-order invariants depending on $\rho,\sigma,\bar{\sigma},\tau,\bar{\tau}$:
\be
f(\rho,\sigma,\bar{\sigma},\tau,\bar{\tau})=
f(\rho',\sigma',\bar{\sigma'},\tau',\bar{\tau'}).
\ee
Using infinitesimal transformations
\be \label{itr}
\lambda= 1+ \check{\lambda},~e^{i\omega}= 1+i\check{\omega},~ \kappa= \check{\kappa}
\ee
with small $\check{\lambda},\check{\omega},\check{\kappa}$ and expanding the rhs 
to first order, we obtain:
\ba
&&\check{\lambda}(\rho\frac{\partial f}{\delta \rho}
+\sigma \frac{\delta f}{\partial \sigma}
+\bar{\sigma} \frac{\partial f}{\partial \bar{\sigma}})
+i\check{\omega}(-2\sigma\frac{\partial f}{\partial \sigma}
+2\bar{\sigma}\frac{\partial f}{\partial \bar{\sigma}}
- \tau\frac{\partial f}{\partial \tau}
+ \bar{\tau}\frac{\partial f}{\partial \bar{\tau}}) \nn \\
&&+\check{\kappa}(-\rho\frac{\partial f}{\partial \tau}
+\bar{\sigma}\frac{\partial f}{\partial \bar{\tau}})
+\check{\bar{\kappa}}(-\rho\frac{\partial f}{\partial \bar{\tau}}
+\sigma\frac{\partial f}{\partial \tau})
+i\bar{\delta}\check{\omega}\frac{\partial f}{\partial \tau}
-i\delta \check{\omega}\frac{\partial f}{\partial \bar{\tau}} =0. \nn
\ea
The relations must be satisfied for all $\check{\lambda},\check{\omega},\check{\kappa}$, leading to 
\be
\frac{\partial f}{\partial \tau}=0,~\frac{\partial f}{\partial \bar{\tau}}=0,~ 
\rho\frac{\partial f}{\partial \rho}+ \sigma\frac{\partial f}{\partial \sigma} +
 \bar{\sigma}\frac{\partial f}{\partial \bar{\sigma}}=0, ~
\sigma\frac{\partial f}{\partial \sigma} -\bar{\sigma}\frac{\partial f}{\partial 
\bar{\sigma}}=0.  
\ee
These are essentially $m=2$ linear partial differential equations for $n=3$
independent variables $\rho,\sigma,\bar{\sigma}$. Systems of differential 
equations of this type are treated, e.g., by Kamke \cite{kamke}. They are 
called involution systems, if the integrability conditions (IC) are satisfied.
With the derivative operators
\be
F^1 = \rho \frac{\partial}{\partial \rho}+\sigma\frac{\partial}{\partial \sigma}
+\bar{\sigma}\frac{\partial}{\partial \bar{\sigma}},~F^2 = \sigma\frac{\partial}{\partial \sigma} - \bar{\sigma}\frac{\partial}{\partial \bar{\sigma}}
\ee
we write differential equations for $f$ 
\be
F^if =0, i=1...m.
\ee
The $m(m-1)/2 =1$ IC's  are then
\be
F^i(F^k)f -F^k(F^i)f =0.          
\ee
In our case the IC's are satisfied. When considering the number of solutions, 
the rank of the coefficient matrix plays a role. This is the $m\times n$ matrix:
\be
\left[ {\begin{array}{ccc}
         \rho & \sigma & \bar{\sigma} \\  
	 0 & \sigma&  -\bar{\sigma}     
\end{array} } \right]
\ee
The rank of this matrix is 2, if $|\sigma| \neq 0$. In this case $n-m=1$ 
function forms an integral basis. Obviously
\be
 j=\rho/|\sigma|
\ee
or any function of j represent the first order invariant. Only this first-order 
invariant exists.

Turning now to second-order invariants, we can write 
\bea \label{icond1}
f(D\rho,D\sigma,D\bar{\sigma},\delta \rho, \bar{\delta} \rho,\delta \sigma,
\bar{\delta} \sigma, \delta \bar{\sigma}, \bar{\delta} \bar{\sigma},
\delta \tau, \bar{\delta} \tau, \delta \bar{\tau}, \bar{\delta} \bar{\tau}, \rho, \sigma, \bar{\sigma}, \tau, \bar{\tau}) = \nn \\
f(D'\rho',D'\sigma',D'\bar{\sigma'},\delta' \rho', \bar{\delta'} \rho',\delta' \sigma',
\bar{\delta'} \sigma', \delta' \bar{\sigma'}, \bar{\delta'} \bar{\sigma'},
\delta' \tau', \bar{\delta'} \tau', \delta' \bar{\tau'}, \bar{\delta'} \bar{\tau'}, \\
\rho', \sigma', \bar{\sigma'}, \tau', \bar{\tau'}).  \nn
\eea
Because of the identity \eref{iiden}, $f$ may be considered as independent of 
$D\tau, D\bar{\tau}$, also the simplified version of the triad is taken.
With the infinitesimal transformations \eref{itr} we have to linear order
\bea
D'\rho' = D\rho +\rho D\check{\lambda}+ 2\check{\lambda}D\rho, \\
D'\sigma' = D\sigma +\sigma D\check{\lambda} -2i\check{\omega}D\sigma + 2\check{\lambda} D\sigma, \\
\delta'\rho' = \delta\rho +\rho\delta\check{\lambda}
+\check{\lambda}\delta\rho + i\check{\omega}\delta\rho 
-\check{\bar{\kappa}}D\rho, \\
\delta'\sigma' = \delta\sigma + \check{\lambda}\delta\sigma 
+\sigma\delta\check{\lambda} -2i\sigma\delta\check{\omega}
-i\check{\omega}\delta\sigma -\check{\bar{\kappa}}D\sigma, \\
\bar{\delta}'\sigma' = \bar{\delta}\sigma + \check{\lambda}\bar{\delta}\sigma 
+\sigma\bar{\delta}\check{\lambda} -2i\sigma\bar{\delta}\check{\omega}
-3i\check{\omega}\bar{\delta}\sigma -\check{\kappa} D\sigma, \\
\delta'\tau' = \delta\tau -i\tau\delta\check{\omega} 
+i\delta\bar{\delta}\check{\omega} + \sigma\delta\check{\bar{\kappa}}
+\check{\bar{\kappa}}\delta\sigma -\rho\delta\check{\kappa}
-\check{\kappa}\delta\rho -\check{\bar{\kappa}}D\tau, \\
\bar{\delta}'\tau' = \bar{\delta}\tau -2i\check{\omega}\bar{\delta}\tau
-i\tau\bar{\delta}\check{\omega} +i\bar{\delta}\bar{\delta}\check{\omega}
+ \sigma\bar{\delta}\check{\bar{\kappa}}+ \check{\bar{\kappa}}\bar{\delta}\sigma
-\rho\bar{\delta}\check{\kappa} -\check{\kappa}\bar{\delta}\rho-
\check{\kappa}D\tau, \\
\rho'= \rho +\check{\lambda}\rho, \\
\sigma' = \sigma +\check{\lambda}\sigma -2i\check{\omega}\sigma, \\
\tau' = \tau -i\check{\omega}\tau + i\bar{\delta}\check{\omega}
+\sigma\check{\bar{\kappa}}- \rho\check{\kappa}.
\eea
Expanding \ref{icond1}, we get
\bea
\check{\lambda}\Bigl(2D\rho\frac{\partial f}{\partial D\rho} +2D\sigma\frac{\partial f}{\partial D\sigma} +2D\bar{\sigma}\frac{\partial f}{\partial D\bar{\sigma}} 
+ \delta\rho\frac{\partial f}{\partial \delta\rho}
+ \bar{\delta}\rho\frac{\partial f}{\partial \bar{\delta}\rho}
+ \delta\sigma\frac{\partial f}{\partial \delta\sigma} \nn \\     
+ \bar{\delta}\sigma \frac{\partial f}{\partial \bar{\delta}\sigma} 
+ \delta\bar{\sigma}\frac{\partial f}{\partial \delta\bar{\sigma}}
+ \bar{\delta}\bar{\sigma}\frac{\partial f}{\partial \bar{\delta}\bar{\sigma}}
+ \rho\frac{\partial f}{\partial \rho} + \sigma\frac{\partial f}{\partial \sigma}
+ \bar{\sigma}\frac{\partial f}{\partial \bar{\sigma}}\Bigr) \nn \\
+ D\check{\lambda}\Bigl(\rho\frac{\partial f}{\partial D\rho}
+ \sigma\frac{\partial f}{\partial D\sigma} 
+\bar{\sigma}\frac{\partial f}{\partial D\bar{\sigma}} \Bigr) \nn \\
+ i\check\omega \Bigl(-2D\sigma\frac{\partial f}{\partial D\sigma}
+ 2D\bar{\sigma}\frac{\partial f}{\partial D\bar{\sigma}}
+ \delta\rho\frac{\partial f}{\partial \delta\rho}
-\bar{\delta}\rho\frac{\partial f}{\partial \bar{\delta}\rho}
-\delta\sigma\frac{\partial f}{\partial \delta\sigma} \nn \\
-3\bar{\delta}\sigma\frac{\partial f}{\partial \bar{\delta}\sigma}
+3\delta\bar{\sigma}\frac{\partial f}{\partial \delta\bar{\sigma}}
+ \bar{\delta}\bar{\sigma}\frac{\partial f}{\partial \bar{\delta}\bar{\sigma}}
+ 2\delta\bar{\tau}\frac{\partial f}{\partial     \delta {\bar\tau}}
- 2\bar{\delta}     \tau \frac{\partial f}{\partial \bar{\delta}     \tau}  \nn \\
-2\sigma\frac{\partial f}{\partial \sigma} 
+2\bar{\sigma}\frac{\partial f}{\partial \bar{\sigma}}
-\tau\frac{\partial f}{\partial \tau} 
+\bar{\tau}\frac{\partial f}{\partial \bar{\tau}} \Bigr) \nn \\
+ i\delta\check{\omega} \Bigl(  -\tau\frac{\partial f}{\partial \delta\tau}
+ \bar{\tau}\frac{\partial f}{\partial      \delta \bar{\tau}}
-\frac{\partial f}{\partial \bar{\tau}}
-2\sigma\frac{\partial f}{\partial \delta\sigma}
+2\bar{\sigma}\frac{\partial f}{\partial \delta\bar{\sigma}} \Bigr) \nn \\
+ i\bar{\delta}\check{\omega}
\Bigl( -\tau\frac{\partial f}{\partial \bar{\delta}\tau}
+\bar{\tau}\frac{\partial f}{\partial \bar{\delta}\bar{\tau}}
+ \frac{\partial f}{\partial \tau}
-2\sigma\frac{\partial f}{\partial \bar{\delta\sigma}}
+2\bar{\sigma}\frac{\partial f}{\partial \bar{\delta}\bar{\sigma}} \Bigr) \nn \\
+ \delta\check{\lambda} \Bigl( \rho\frac{\partial f}{\partial \delta\rho}
+ \sigma\frac{\partial f}{\partial \delta\sigma}
+ \bar{\sigma}\frac{\partial f}{\partial \delta\bar{\sigma}} \Bigr) 
+ \bar{\delta}\check{\lambda} 
\Bigl( \rho\frac{\partial f}{\partial \bar{\delta}\rho}
+ \sigma\frac{\partial f}{\partial \bar{\delta} \sigma}
+ \bar{\sigma}\frac{\partial f}{\partial \bar{\delta}\bar{\sigma}} \Bigr) \nn \\
+ \check{\bar{\kappa}} \Bigl(-D\rho\frac{\partial f}{\partial \delta\rho}
-D\sigma\frac{\partial f}{\partial \delta\sigma}
-D\bar{\sigma}\frac{\partial f}{\partial \delta\bar{\sigma}}
+\delta\sigma\frac{\partial f}{\partial \delta\tau}
-D\tau\frac{\partial f}{\partial \delta\tau}
-\delta\rho\frac{\partial f}{\partial      \delta\bar{\tau}} \nn \\
-D\bar{\tau}\frac{\partial f}{\partial     \delta\bar{\tau}}
+\bar{\delta}\sigma\frac{\partial f}{\partial \bar{\delta}\tau}
-\bar{\delta}\rho \frac{\partial f}{\partial \bar{\delta}\bar{\tau}}
+\sigma\frac{\partial f}{\partial \tau}
-\rho\frac{\partial f}{\partial \bar{\tau}} \Bigr) \nn \\
+ \check{\kappa} \Bigl(-D\rho\frac{\partial f}{\partial \bar{\delta}\rho}
-D\sigma\frac{\partial f}{\partial \bar{\delta}\sigma}
-D\bar{\sigma}\frac{\partial f}{\partial \bar{\delta}\bar{\sigma}}
   -\delta\rho\frac{\partial f}{\partial \delta\tau}
    -D\bar{\tau}\frac{\partial f}{\partial \bar{\delta}\bar{\tau}}
   + \delta\bar{\sigma}\frac{\partial f}{\partial      \delta \bar{\tau}} \nn \\
   -D\tau\frac{\partial f}{\partial \bar{\delta}     \tau} 
   +\bar{\delta}\bar{\sigma}\frac{\partial f}{\partial \bar{\delta}\bar{\tau}}
   -\bar{\delta}\rho \frac{\partial f}{\partial \bar{\delta}    \tau} 
  -\rho\frac{\partial f}{\partial \tau}
  +\bar{\sigma}\frac{\partial f}{\partial \bar{\tau}} \Bigr) \nn \\
+ \delta\check{\bar{\kappa}} \Bigl(
\sigma\frac{\partial f}{\partial \delta\tau}
-\rho \frac{\partial f}{\partial \delta\bar{\tau}} \Bigr) 
+ \delta\check{\kappa} 
\Bigl(\bar{\sigma}\frac{\partial f}{\partial \delta\bar{\tau}}
- \rho\frac{\partial f}{\partial \delta\tau} \Bigr) \nn \\
+ \bar{\delta}\check{\kappa} \Bigl(
- \rho\frac{\partial f}{\partial \bar{\delta}\tau}
+ \bar{\sigma}\frac{\partial f}{\partial \bar{\delta}\bar{\tau}} \Bigr)
+ \bar{\delta}\check{\bar{\kappa}} \Bigl(
\sigma \frac{\partial f}{\partial \bar{\delta}\tau}
-\rho \frac{\partial f}{\partial \bar{\delta} \bar{\tau}} \Bigr) \nn \\
-i\delta\delta\check{\omega} \frac{\partial f}{\partial \delta \bar{\tau}}
+i\bar{\delta}\bar{\delta}\check{\omega}\frac{\partial f}{\partial
\bar{\delta}\tau} 
+ i\delta\bar{\delta}\check{\omega} \frac{\partial f}{\partial \delta\tau}
- i\bar{\delta}\delta\check{\omega} \frac{\partial f}{\partial \bar{\delta}\bar{\tau}}.   \nn
\eea

The equation must be satisfied for any values of $\check{\kappa},\check{\lambda},\check{\omega}$, so that we obtain 
\bea
2D\rho\frac{\partial f}{\partial D\rho} 
+2D\sigma\frac{\partial f}{\partial D\sigma}
+2D\bar{\sigma}\frac{\partial f}{\partial D\bar{\sigma}} 
+ \delta\rho\frac{\partial f}{\partial \delta\rho}
+ \bar{\delta}\rho\frac{\partial f}{\partial \bar{\delta}\rho}
+ \delta\sigma\frac{\partial f}{\partial \delta\sigma} \nn \\     
+ \bar{\delta}\sigma \frac{\partial f}{\partial \bar{\delta}\sigma} 
+ \delta\bar{\sigma}\frac{\partial f}{\partial \delta\bar{\sigma}}
+ \bar{\delta}\bar{\sigma}\frac{\partial f}{\partial \bar{\delta}\bar{\sigma}}
+ \rho\frac{\partial f}{\partial \rho} + \sigma\frac{\partial f}{\partial \sigma}
+ \bar{\sigma}\frac{\partial f}{\partial \bar{\sigma}}=0,  \\
\rho\frac{\partial f}{\partial D\rho}
+ \sigma\frac{\partial f}{\partial D\sigma} 
+ \bar{\sigma}\frac{\partial f}{\partial D\bar{\sigma}} =0,  \\
 -2D\sigma\frac{\partial f}{\partial D\sigma}
+ 2D\bar{\sigma}\frac{\partial f}{\partial D\bar{\sigma}}
+ \delta\rho\frac{\partial f}{\partial \delta\rho}
-\bar{\delta}\rho\frac{\partial f}{\partial \bar{\delta}\rho}
-\delta\sigma\frac{\partial f}{\partial \delta\sigma}  
-3\bar{\delta}\sigma\frac{\partial f}{\partial \bar{\delta}\sigma} \nn \\
+3\delta\bar{\sigma}\frac{\partial f}{\partial \delta\bar{\sigma}}
+ \bar{\delta}\bar{\sigma}\frac{\partial f}{\partial \bar{\delta}\bar{\sigma}}
-2\sigma\frac{\partial f}{\partial \sigma} 
+2\bar{\sigma}\frac{\partial f}{\partial \bar{\sigma}} 
-\tau\frac{\partial f}{\partial \tau} 
+\bar{\tau}\frac{\partial f}{\partial \bar{\tau}}=0,  \\
-\frac{\partial f}{\partial \bar{\tau}}
-2\sigma\frac{\partial f}{\partial \delta\sigma}
+2\bar{\sigma}\frac{\partial f}{\partial \delta\bar{\sigma}}=0, \\          
-\frac{\partial f}{\partial \tau}
+2\sigma\frac{\partial f}{\partial \bar{\delta}\sigma}
-2\bar{\sigma}\frac{\partial f}{\partial \bar{\delta}\bar{\sigma}}=0, \\          
 \rho\frac{\partial f}{\partial \delta\rho}
+ \sigma\frac{\partial f}{\partial \delta\sigma}
+ \bar{\sigma}\frac{\partial f}{\partial \delta\bar{\sigma}}=0, \\
 \rho\frac{\partial f}{\partial \bar{\delta}\rho}
+ \sigma\frac{\partial f}{\partial \bar{\delta} \sigma}
+ \bar{\sigma}\frac{\partial f}{\partial \bar{\delta}\bar{\sigma}} =0,\\
 D\rho\frac{\partial f}{\partial \delta\rho}
+D\sigma\frac{\partial f}{\partial \delta\sigma}
+D\bar{\sigma}\frac{\partial f}{\partial \delta\bar{\sigma}}
-\sigma\frac{\partial f}{\partial \tau}
+\rho\frac{\partial f}{\partial \bar{\tau}} = 0, \\
 D\rho\frac{\partial f}{\partial \bar{\delta}\rho}
+D\sigma\frac{\partial f}{\partial \bar{\delta}\sigma}
+D\bar{\sigma}\frac{\partial f}{\partial \bar{\delta}\bar{\sigma}}
  +\rho\frac{\partial f}{\partial \tau}
  -\bar{\sigma}\frac{\partial f}{\partial \bar{\tau}} =0, \\
\frac{\partial f}{\partial \bar{\delta}\tau}= 0,~
\frac{\partial f}{\partial \delta\bar{\tau}}= 0,~
\frac{\partial f}{\partial \delta\tau}= 0,~
\frac{\partial f}{\partial \bar{\delta}\bar{\tau}}=0.  
\eea
Let us first assume $\rho\neq 0,|\sigma|\neq 0$. We have 9 partial differential 
equations $F^if=0$ for $f$, depending on 14 independent variables 
$D\rho,D\sigma,D\bar{\sigma},\delta\rho,\bar{\delta}\rho,\delta\sigma,\bar{\delta}\sigma,\delta\bar{\sigma},\bar{\delta}\bar{\sigma},\rho,\sigma,\bar{\sigma},\tau, \bar{\tau}.$ 
The 36 integrability conditions $F^i(F^k)f-F^k(F^i)f =0$ are satisfied: They are identical to $F^if=0$, as shown by a direct calculation. 
The system can be resolved with respect to the first $m=9$ derivatives of $f$, 
$\frac{\delta f}{D\rho}, ...\frac{\delta f}{\bar{\delta}\bar{\sigma}}$. 
One thus obtains its so-called canonical form.  However, it should be noted that 
some denominators may become zero in this process, meaning that singular cases 
must be considered. All resolved derivatives contain the factor 
\be
\rho|\sigma|I_2
\ee
in their denominators. Here $I_2$ is the imaginary part of the complex quantity 
\be
I=I_1+iI_2 = \frac{i}{|\sigma|}\Bigl(\frac{D\rho}{\rho} -\frac{D\sigma}{\sigma}\Bigr)+2\frac{\nu}{|\sigma|}.
\ee
$I$ is itself an invariant of the system, written here with a non-vanishing 
rotation coefficient $\nu$. To simplify the calculations, we assumed $\nu=0$.
In the invariants $J,L,M,N$ to be derived, $\nu$ only appears via $I$, thus
no transformation to $\nu \neq 0$ is necessary.

Obviously, we have to consider the following cases separately:
\bea
 |\sigma| \neq 0,~\rho \neq 0,~I_2 \neq 0, \\
 |\sigma| \neq 0,~\rho \neq 0,~I_2= 0, \\
 |\sigma| \neq 0,~\rho = 0,  \\
 |\sigma| =0, ~\rho = 0.
\eea
We start the first case

\underline{$|\sigma| \neq 0,~\rho \neq 0,~I_2 \neq 0$}

The invariants $I_1, I_2, j$ may be used to find further invariants,
following a method described by Kamke \cite{kamke}, p. 54. 
The independent variables 
$D\sigma,~D\bar{\sigma},~\rho$ can be replaced by the invariants $I_1,I_2,j$:
\bea
D\sigma = (iI_1-I_2)\sigma\sqrt{\sigma\bar{\sigma}} 
  +D\rho \sqrt{\sigma}/(\sqrt{\bar{\sigma}}j), \\
D\bar{\sigma} =-(iI_1+I_2)\bar{\sigma}\sqrt{\sigma\bar{\sigma}} 
  +D\rho \sqrt{\bar{\sigma}}/(\sqrt{\sigma}j),  \\ 
\rho = j \sqrt{\sigma \bar{\sigma}}.
\eea
The function $f$ therefore no longer depends on 
$D\sigma,D\bar{\sigma},\rho$. $I_1,I_2,j$ are to be regarded as parameters.
We have to consider the new system
\bea
  \sigma\frac{\partial f}{\partial \sigma}
  + \bar{\sigma}\frac{\partial f}{\partial \bar{\sigma}}  
  + \delta\rho\frac{\partial f}{\partial \delta\rho}
  + \bar{\delta}\rho\frac{\partial f}{\partial \bar{\delta}\rho}
  + \delta\sigma\frac{\partial f}{\partial \delta\sigma} \nn \\     
  + \bar{\delta}\sigma \frac{\partial f}{\partial \bar{\delta}\sigma} 
  + \delta\bar{\sigma}\frac{\partial f}{\partial \delta\bar{\sigma}}
  + \bar{\delta}\bar{\sigma}\frac{\partial f}{\partial \bar{\delta}\bar{\sigma}}
  =0,  \\
 \frac{\partial f}{\partial D\rho} =0, \\
   \delta\rho\frac{\partial f}{\partial \delta\rho}
  -\bar{\delta}\rho\frac{\partial f}{\partial \bar{\delta}\rho}
  -\delta\sigma\frac{\partial f}{\partial \delta\sigma}  
  -3\bar{\delta}\sigma\frac{\partial f}{\partial \bar{\delta}\sigma} \nn \\
  +3\delta\bar{\sigma}\frac{\partial f}{\partial \delta\bar{\sigma}}
  +\bar{\delta}\bar{\sigma}\frac{\partial f}{\partial \bar{\delta}\bar{\sigma}}
     -2\sigma\frac{\partial f}{\partial \sigma} 
     +2\bar{\sigma}\frac{\partial f}{\partial \bar{\sigma}} 
     -\tau\frac{\partial f}{\partial \tau} 
     +\bar{\tau}\frac{\partial f}{\partial \bar{\tau}}
      =0,  \\
-\frac{\partial f}{\partial \bar{\tau}}
-2\sigma\frac{\partial f}{\partial \delta\sigma}
+2\bar{\sigma}\frac{\partial f}{\partial \delta\bar{\sigma}}=0, \\ 
-\frac{\partial f}{\partial \tau}
+2\sigma\frac{\partial f}{\partial \bar{\delta}\sigma}
-2\bar{\sigma}\frac{\partial f}{\partial \bar{\delta}\bar{\sigma}}=0, \\ 
 j|\sigma|\frac{\partial f}{\partial \delta\rho}
+ \sigma\frac{\partial f}{\partial \delta\sigma}
+ \bar{\sigma}\frac{\partial f}{\partial \delta\bar{\sigma}}=0, \\
 j|\sigma|\frac{\partial f}{\partial \bar{\delta}\rho}
+ \sigma\frac{\partial f}{\partial \bar{\delta} \sigma}
+ \bar{\sigma}\frac{\partial f}{\partial \bar{\delta}\bar{\sigma}} =0,\\
  j^2|\sigma|^2 \frac{\partial f}{\partial \bar{\tau}}
  -j|\sigma|^2 e^{is} \frac{\partial f}{\partial \tau}
 + j|\sigma| D\rho\frac{\partial f}{\partial \delta\rho}  \nn \\
 + \sigma(ij|\sigma|^2I_1 -j|\sigma|^2I_2  +D\rho)
   \frac{\partial f}{\partial \delta\sigma} \nn \\
 + \bar{\sigma}(-ij|\sigma|^2I_1- j|\sigma|^2I_2 +D\rho)
   \frac{\partial f}{\partial \delta\bar{\sigma}} =0, \\
  j^2|\sigma|^2 \frac{\partial f}{\partial \tau}
  -j|\sigma|^2 e^{-is} \frac{\partial f}{\partial \bar{\tau}}
 + j|\sigma| D\rho\frac{\partial f}{\partial \bar{\delta}\rho} \nn \\
 + \sigma(ij|\sigma|^2I_1 -j|\sigma|^2I_2  +D\rho)
   \frac{\partial f}{\partial \bar{\delta}\sigma} \nn \\
 + \bar{\sigma}(-ij|\sigma|^2I_1- j|\sigma|^2I_2 +D\rho)
   \frac{\partial f}{\partial \bar{\delta}\bar{\sigma}} =0
\eea
for the new function $f=f(D\rho,\delta\rho,\bar{\delta}\rho,\delta\sigma,\bar{\delta}\sigma,\delta\bar{\sigma},\bar{\delta}\bar{\sigma},\sigma,\bar{\sigma},\tau, \bar{\tau}).$ 
   The second equation shows that $f$ is also independent of $D\rho$.
The system then consists of 8 equations for 10 independent variables 
$\delta\rho...\bar{\tau}$. An integral basis should therefore consist of 10-8=2 functions. The complex quantity 
\bea
J=J_1+iJ_2 = e^{is/2}\Bigl[(\vartheta\bar{\vartheta}-4)\Bigl(\frac{\delta\sigma}{\sigma}
  -\frac{\delta\rho}{\rho}\Bigr)+(\zeta\bar{\vartheta}-4)\Bigl(\frac{\delta\bar{\sigma}}{\bar{\sigma}}
  -\frac{\delta\rho}{\rho}\Bigr) \nn \\
  +2\bar{\vartheta}(\zeta-\vartheta)\bar{\tau}\Bigr]
  + 2e^{-is/2}(\zeta-\vartheta)\Bigl(\frac{\bar{\delta}\sigma}{\sigma}
  -\frac{\bar{\delta}\rho}{\rho} +2\tau\Bigr)
\eea
with
\be
\zeta=2j-iI_1+I_2,~\vartheta=2j-iI_1-I_2,~e^{is}= \sigma/|\sigma|
\ee
is a solution and represents an integral basis.- 
In the case 

\underline{$|\sigma| \neq 0,~\rho \neq 0,~ I_2=0$},

the invariant $J$ simplifies to
\be
J = (I_1^2-4(1-j^2))M,
\ee
where 
\be
M =e^{is/2}(\frac{\delta\sigma}{\sigma}
   +\frac{\delta\bar{\sigma}}{\bar{\sigma}}-2\frac{\delta\rho}{\rho})
\ee 
itself is a new complex invariant in the case $I_2=0$. As abbreviation we frequently use
\be
I_3= I_1^2 -4(1-j^2).
\ee
If also $I_3=0$ applies, there is a further invariant
\be L = e^{is/2} (\frac{\delta\sigma}{\sigma}-\frac{\delta\rho}{\rho}
-2\bar{\tau}) + e^{-is/2}(j+i\sqrt{1-j^2})  (\frac{\bar{\delta}\bar{\sigma}}{\bar{\sigma}} -\frac{\bar{\delta}\rho}{\rho} -2\tau).     
\ee
The next special case to consider is non-vanishing shear with 
vanishing divergence:

\underline{$|\sigma| \neq 0,~\rho =0$} 

Here the following system must be solved;
\bea
 2D\sigma\frac{\partial f}{\partial D\sigma}
+2D\bar{\sigma}\frac{\partial f}{\partial D\bar{\sigma}} 
+ \delta\sigma\frac{\partial f}{\partial \delta\sigma} \nn \\     
+ \bar{\delta}\sigma \frac{\partial f}{\partial \bar{\delta}\sigma} 
+ \delta\bar{\sigma}\frac{\partial f}{\partial \delta\bar{\sigma}}
+ \bar{\delta}\bar{\sigma}\frac{\partial f}{\partial \bar{\delta}\bar{\sigma}}
+ \sigma\frac{\partial f}{\partial \sigma}
+ \bar{\sigma}\frac{\partial f}{\partial \bar{\sigma}}=0,  \\
 \sigma\frac{\partial f}{\partial D\sigma} 
+ \bar{\sigma}\frac{\partial f}{\partial D\bar{\sigma}} =0,  \\
 -2D\sigma\frac{\partial f}{\partial D\sigma}
+ 2D\bar{\sigma}\frac{\partial f}{\partial D\bar{\sigma}}
-\delta\sigma\frac{\partial f}{\partial \delta\sigma}  
-3\bar{\delta}\sigma\frac{\partial f}{\partial \bar{\delta}\sigma} \nn \\
+3\delta\bar{\sigma}\frac{\partial f}{\partial \delta\bar{\sigma}}
+ \bar{\delta}\bar{\sigma}\frac{\partial f}{\partial \bar{\delta}\bar{\sigma}}
-2\sigma\frac{\partial f}{\partial \sigma} 
+2\bar{\sigma}\frac{\partial f}{\partial \bar{\sigma}} 
-\tau\frac{\partial f}{\partial \tau} 
+\bar{\tau}\frac{\partial f}{\partial \bar{\tau}}=0,  \\
-\frac{\partial f}{\partial \bar{\tau}}
-2\sigma\frac{\partial f}{\partial \delta\sigma}
+2\bar{\sigma}\frac{\partial f}{\partial \delta\bar{\sigma}}=0, \\          
-\frac{\partial f}{\partial \tau}
+2\sigma\frac{\partial f}{\partial \bar{\delta}\sigma}
-2\bar{\sigma}\frac{\partial f}{\partial \bar{\delta}\bar{\sigma}}=0, \\        
  \sigma\frac{\partial f}{\partial \delta\sigma}
+ \bar{\sigma}\frac{\partial f}{\partial \delta\bar{\sigma}}=0, \\
  \sigma\frac{\partial f}{\partial \bar{\delta} \sigma}
+ \bar{\sigma}\frac{\partial f}{\partial \bar{\delta}\bar{\sigma}} =0,\\
 D\sigma\frac{\partial f}{\partial \delta\sigma}
+D\bar{\sigma}\frac{\partial f}{\partial \delta\bar{\sigma}}
-\sigma\frac{\partial f}{\partial \tau} = 0, \\
 D\sigma\frac{\partial f}{\partial \bar{\delta}\sigma}+
 D\bar{\sigma}\frac{\partial f}{\partial \bar{\delta}\bar{\sigma}}
  -\bar{\sigma}\frac{\partial f}{\partial \bar{\tau}} =0, \\
\frac{\partial f}{\partial \bar{\delta}\tau}= 0,~
\frac{\partial f}{\partial \delta\bar{\tau}}= 0,~
\frac{\partial f}{\partial \delta\tau}=
\frac{\partial f}{\partial \bar{\delta}\bar{\tau}}.  
\eea
This time, there are 9 differential equations for 10 independent variables 
$D\sigma,D\bar{\sigma},\delta\sigma,\bar{\delta}\sigma,\delta\bar{\sigma},\bar{\delta}\bar{\sigma},\sigma,\bar{\sigma},\tau,\bar{\tau}$.  
 The integrability conditions are also satisfied, as can be easily verified.
When attempting to convert the equations into canonical form, again terms appear 
in some denominators that become zero in the special case $I_1=0$, 
equivalent to $Ds=0$. We ignore this case initially.
The last equation in the canonical version can be written: 
\be
\frac{\delta f}{\delta \bar{\delta}\bar{\sigma}}(4-I_1^2)
\frac{|\sigma|^3e^{-is}}{Ds} =0.
\ee
In the case of $I_1^2=4$, this equation drops out, and the integration of the remaining equations yields the invariant
\bea
N= e^{is/2}(\frac{\delta \sigma}{\sigma}-\frac{\delta \bar{\sigma}}{\bar{\sigma}}-4\bar{\tau})/(1+i)
-ie^{-is/2}( \frac{\bar{\delta} \sigma}{\sigma}
 -\frac{\bar{\delta} \bar{\sigma} }{\bar{\sigma}} +4\tau)/(1+i).
\eea
The apparently complex function $N$ is actually real and represents one single 
invariant. If $I_1^2 \neq 4$, the only invariant for this case is 
- apart from $j=0$ - $I_1$. This case also includes the special case $I_1=0$ above.

In the case
\underline{$|\sigma|=0,~\rho \neq 0$}
we are looking for invariants with dependencies 
$f=f(D\rho,\delta \rho, \bar{\delta} \rho,
\delta \tau, \bar{\delta} \tau, \delta \bar{\tau}, 
\bar{\delta} \bar{\tau}, \rho,  \tau, \bar{\tau})$.
Using the same methods as before, it can be shown that no invariants exist.

The last case is
\underline{$|\sigma|=0,~\rho=0$}.
The invariants depend here on the independent variables $\delta\tau, 
   \bar{\delta}\tau, \delta\bar{\tau}, \bar{\delta}\bar{\tau}, \tau,\bar{\tau}$. 
One obtains the differential equations 
\bea
2\tau\frac{\partial f}{\partial \delta\tau}
    +\frac{\partial f}{\partial \bar{\tau}} =0,~
\bar{\tau}(\frac{\partial f}{\partial \delta\tau}+
	   \frac{\partial f}{\partial \bar{\delta}\bar{\tau}})
	   +\frac{\partial f}{\partial \tau} =0,~
-\tau\frac{\partial f}{\partial \tau}
   +\bar{\tau}\frac{\partial f}{\partial \bar{\tau}} =0,~\\
   \frac{\partial f}{\partial \delta\bar{\tau}} =
   \frac{\partial f}{\partial \bar{\delta}\tau} =0,~
\frac{\partial f}{\partial \delta\tau}=
\frac{\partial f}{\partial \bar{\delta}\bar{\tau}}. 
\eea
A solution $f$ is the Gaussian curvature 
\be \nn 
K= -2\tau\bar{\tau} + \bar{\delta}\bar{\tau} +\delta\tau.
\ee
or any function of $K$.

Appendix $A$ provides a brief overview of the results of this section.
One important point should be mentioned: 
\emph{There exist no unique set of invariants valid for all null 
hypersurfaces}. Different classes of hypersurfaces admit different sets of 
differential invariants.     


\section{The intrinsic Killing equation for $\mathcal{N}_3$.}                 

After these preparations we discuss Killing isometries of null hypersurfaces.  
A $\mathcal{N}_3$ admits an inner isometry group if the Lie derivative of the inner metric $\gamma_{ik}$ vanishes with respect to a Killing field $\xi^i(x^i)$. The Lie derivative is already defined on a differentiable manifold and therefore independent of any affinity given additionally. It can be applied to degenerate metrics such as that of a null hypersurface, where an affinity is not uniquely determined:
\be \label{eq0}
\stackrel[\xi^i]{}{L}\gamma_{ik} \equiv \frac{\partial\gamma_{ik}}{\partial x^l}\xi^l +\gamma_{kl}\frac{\partial \xi^l}{\partial x^i} +\gamma_{il}\frac{\partial \xi^l}{\partial x^k}.
  \ee
But with the affinities \eref{aff} the Lie derivative may also be written
\be \label{0k1}
  \stackrel[\xi^i]{}{L}\gamma_{ik} = \xi^l\nabla_l\gamma_{ik} 
  +\gamma_{kl}\nabla_i\xi^l +\gamma_{il}\nabla_k\xi^l,
  \ee
  where $\nabla_i$ is the covariant derivative associated with the affinity. With $\xi_k=\gamma_{kl}\xi^l$ and $\nabla_i\gamma_{kl} = \gamma_kh_{li} +\gamma_lh_{ki}$ the expression is transformed into a third form
  \be
  \stackrel[\xi^i]{}{L}\gamma_{ik} = \nabla_k\xi_i +\nabla_i\xi_k -2 \xi^l\gamma_lh_{ik}.
  \ee
   $h_{ik}$ is the Lie derivative of $\gamma_{ik}$ with regard to $\epsilon^l$:
   \be \label{eq1}
   h_{ik}=  \stackrel[\epsilon^i]{}{L}\gamma_{ik}= \frac{1}{2}(\partial_k\gamma_{il} + \partial_i\gamma_{kl} - \partial_l\gamma_{ik})\epsilon^l,
   \ee
where $\partial_k$ is the ordinary derivative. 
Thus the intrinsic Killing equation (KE) of a null hypersurface can also be written in the form
   \be  \label{keq}         
   \nabla_k\xi_l +\nabla_i\xi_k = 2 h_{ik}\xi^l\gamma_l,
   \ee 
which differs from the usual covariant KE in non-degenerated metrics by an addtional term on the rhs.

We now ask which groups of motions on null hypersurfaces exist and which metrics are compatible with the existence of motions. The discussion is limited to local areas, where a Killing vector (KV) $\xi^i$ points either everywhere or nowhere in the direction of the generators. Likewise, the assumptions made below about the disappearance or non-disappearance of invariants or covariants shall always apply to a finite subdomain.

In this section, the \emph{triad formalism} is tentatively used to answer the questions. In general, we can represent $\xi^i$  as
\be \label{0ka}
\xi^i = \xi t^i +\bar{\xi}\bar{t}^i + \eta\epsilon^i.
\ee
The KE can then be written
\ba\label{0k2}
\bar{\delta}\xi +\xi\tau -\eta\sigma &=& 0,  \\  \label{0k3}
\delta\xi + \bar{\delta}\bar{\xi} -\xi \bar{\tau}- \bar{\xi}\tau
-2\rho\eta &=& 0,  \\
\label{0k4}
D\xi +(\rho+i\nu)\xi +\sigma\bar{\xi} &=& 0.
\ea
Obviously there exist two types of solutions of the KE: $\xi^i$ can point in the generator (null) direction (hence $\xi=0$) 
or in space-like directions ($\xi\bar{\xi} \neq 0$). 
In each case we have to study the integrability conditions of the KE.
Their discussion requires some case distinctions.  
We assume that the conditions for each case are always valid for a finite 
domain and start with the general case 
$\rho \neq 0,~ \sigma \neq 0,~ I_2 \neq 0$.
\eref{0k3} permits to eliminate  $\eta$: 
\be
\label{0k5}
\eta = \frac{1}{2\rho}(\delta \xi +\bar{\delta}\bar{\xi}-\xi\bar{\tau}
- \bar{\xi}\tau).
\ee
\eref{0k2} then takes the form
\be
\label{0k6}
\bar{\delta}\xi + \xi\tau - \frac{\sigma}{2\rho}(\delta\xi 
+\bar{\delta}\bar{\xi} -\xi\bar{\tau} -\bar{\xi}\tau)=0.
\ee
If we specify an $\xi$ in accordance with \eref{0k6} on an initial two-dimensional space-like surface, \eref{0k4} allows the determination of $\xi$ in a finite domain by integration along the null direction. When progressing in the null direction, \eref{0k6} is generally not preserved. 
We therefore have to demand $D\eref{0k6} = 0$. This gives, if the communicator relations \eref{commu1}-\eref{commu3} are observed: 
\be \label{0k7}
\delta\xi(1- \frac{iI}{2j}) -\bar{\delta}\bar{\xi}(1+\frac{iI}{2j})=
A\xi +B\bar{\xi},
\ee  
\ba
A = \frac{\delta\sigma}{\sigma}-\frac{\delta\rho}{\rho}
- \bar{\tau}(1+\frac{iI}{2j}), ~ B = \frac{\bar{\delta}\sigma}{\sigma}- 
\frac{\bar{\delta}\rho}{\rho} + \tau(1-\frac{iI}{2j}).
\ea
Equation \eref{0k7} is a system of algebraic equations for $\delta\xi, \bar{\delta}\bar{\xi}$. If the determinant
\be
 \frac{i}{j}(\bar{I}-I) = \frac{2I_2}{j} \neq 0,
\ee
i.e. if the invariant $I_2$ is  different from zero as supposed, the solution is
\be \label{0k8}
\delta\xi = R\xi +S\bar{\xi}
\ee
with
\ba
R &=& \frac{j}{2I_2}(\frac{\delta\sigma}{\sigma}-\frac{\delta\rho}{\rho})
(1+\frac{i\bar{I}}{2j})
+\frac{j}{2I_2}(\frac{\delta\bar{\sigma}}{\bar{\sigma}}-\frac{\delta\rho}{\rho})
(1+\frac{iI}{2j}),\\
S &=& \frac{j}{2I_2}(\frac{\bar{\delta}\sigma}{\sigma}
-\frac{\bar{\delta}\rho}{\rho}) (1+\frac{i\bar{I}}{2j})
+\frac{j}{2I_2}(\frac{\bar{\delta}\bar{\sigma}}{\bar{\sigma}}
-\frac{\bar{\delta}\rho}{\rho}) (1+\frac{iI}{2j}) +\tau.
\ea

Altogether, instead of \eref{0k2}-\eref{0k4}, we get the system \eref{0k8}
and 
\ba  \label{0k9}
\bar{\delta}\xi &=& \frac{\sigma\xi}{2\rho}(R+\bar{S} -\bar{\tau})- \xi\tau 
+\frac{\sigma\bar{\xi}}{2\rho}(S +\bar{R}-\tau),  \\ \label{0k10}
D\xi &=& -(\rho+i\nu)\xi -\sigma\bar{\xi} 
\ea
We have to add the equation giving the Killing vector component in the null direction, $\eta$:
\be \label{0k12}
\eta = \frac{1}{2\rho}\xi(R+\bar{S} -\bar{\tau}) 
+\frac{1}{2\rho}\bar{\xi}(\bar{R}+S -\tau).
\ee
The equations \eref{0k8},\eref{0k9},\eref{0k10} are a system of linear differential equations for the complex function $\xi$. 
For such systems a well-known theorem applies \cite{eisenhart1}, shortly "Eisenharts theorem":	
The system is completely integrable, if the integrability conditions are satisfied identically. One obtains a first set of integrability conditions from the commutator relations \eref{commu1}-\eref{commu3} applied to $\xi$:
\be \label{kom1}
D\delta\xi-\delta D\xi  =  (\rho +i\upsilon)\delta\xi +\bar{\sigma}\bar{\delta}\xi,                    
\ee
\be \label{kom2}
D\bar{\delta}\xi-\bar{\delta} D\xi  =
(\rho -i\upsilon)\bar{\delta}\xi +\sigma\delta\xi,
\ee
\be \label{kom3}
\delta\bar{\delta}\xi-\bar{\delta}\delta\xi = \bar{\tau}\bar{\delta}\xi-\tau\delta\xi.
 \ee
If we substitute the derivatives of $\xi$,  \eref{kom1} leads to
\be \label{0k13}
\xi L_1 + \bar{\xi}M_1 =0
\ee

with
\ba
L_1 &=& -\delta(\rho+i\nu) -DR +\bar{\sigma}S +(\rho+i\upsilon)R- \bar{\sigma}\tau,\\
M_1 &=& -\delta\sigma -DS +\sigma R +(\rho-i\upsilon)S+ \sigma\bar{\tau}. 
\ea
\eref{0k13} is a homogeneous system of algebraic equations for $\xi, \bar{\xi}$. 
It has the algebraic solvability condition 
\be \label{0k14}
                 L_1\bar{L}_1 -M_1\bar{M}_1  =0.
\ee
This condition is necessary for the existence of a Killing symmetry on $\mathcal{N}_3$  
in the general case $\rho \neq 0,~\sigma \neq 0, I_2 \neq 0$. 
Suppose it holds, we have to distinguish three cases. First, let
\begin{displaymath} \label{0k15}
 M_1\bar{M}_1+L_1\bar{L}_1 \neq 0, ~ M_1+\bar{L}_1 \neq 0. 
\end{displaymath}
The solution in this case is
\be \label{0k16}
\xi = iw(\bar{L}_1+M_1)
\ee
with real $w$. 
In the case
\begin{displaymath}
~ M_1\bar{M}_1+L_1\bar{L}_1  \neq 0, ~M_1+\bar{L}_1  =0
\end{displaymath}
the solution can be written \be
\xi = w M_1
\ee
again with real $w$. In the third case $L_1 = M_1 = 0$
the inegrability conditions are satisfied without restriction for $\xi$. 

The two other equations \eref{kom2}, \eref{kom3} lead to similar homogeneous algebraic equations for $\xi$, with still more complicated invariant coefficients $L_2, M_2, M_3, M_3$, and to additional necessary conditions for the existence of a spacelike Killing symmetry:
\be
L_2\bar{L}_2 -M_2\bar{M}_2 =0,~ L_3\bar{L}_3 -M_3\bar{M}_3 =0.
\ee
Therefore, in the general case $\rho \neq 0,~ \sigma \neq 0,~ I_2 \neq 0$ the integrability conditions split into a large number of cases.
They are difficult to handle without an elaborated theory of higher-order null hypersurface invariants, which does not yet exist. We will therefore use the \emph{coordinate formalism} for further calculations. Special cases like null Killing vectors can still be treated with the triad formalism.

\section{Groups of motion with a null Killing vector}

\subsection{Basic equations}
Consider the case that a Killing vector $\xxi{}^i$ 
points to the null direction $\epsilon^i$, hence $\xi =0$ in the representation  \eref{0k1}. From the Killing equations \eref{0k2}-\eref{0k4} we conclude that the divergence $\rho$ and shear $\sigma$ vanish. Equivalently, the Lie derivative of the metric with regard to $\epsilon^i$ vanishes, $h_{ik}=0$.

This type of null hypersurface is called Killing horizon (or simply horizon).
Physically there are many different types of horizons, depending on their embedding in ambient spacetime. But from the point of view of inner geometry  
they all reduce to a null hypersurface with vanishing shear and divergence.

The Lie group of a horizon represents an infinite Lie group $G_\infty$ rather than a one-dimensional group, since with $X=\xi^k\partial_k$ every $X' = f(x^i)X$  is also a group generator, $f(x^i) \neq 0$ is an otherwise arbitrary function of position. The metric of a general Killing horizon is in adapted coordinates:
\be \label{hor1}
\bold{G_\infty}:~~~ E=E(x_2,x_3),~F=F(x_2,x_3),~G=G(x_2,x_3).
\ee
This is the metric of a general two-dimensional surface. Its Gaussian
curvature $K$ (see \eref{kkk}) is the only second-order differential invariant of horizons.

Horizons may admit additionally motions with spacelike trajectories. 
The KEs for horizons simplify to
\ba
\bar{\delta}\xi +\xi\tau &=& 0,  \label{kke1}\\
\delta\xi + \bar{\delta}\bar{\xi} -\xi \bar{\tau}- \bar{\xi}\tau &=& 0, \label{kke2} \\ 
D\xi +i \nu\xi &=& 0.\label{kke3}
\ea
Obviously, $\xi =0$ is the horizon solution. We look for additional solution with $|\xi\bar{\xi}| \neq 0$.

In order to apply Eisenhart's theorem, 
the form of \eref{kke1}-\eref{kke3} is not yet suitable: 
Not all first-order derivatives 
can be expressed by the functions themselves.
 However, this is the case for the equivalent system:
\ba
\delta h &=& \xi \delta\bar{\tau}+ \bar{\tau}h +K \bar{\xi}, \label{kkee1} \\
\bar{\delta}h &=& \xi(\tau\bar{\tau}-\delta\tau),\label{kkee2}   \\
Dh &=& -i\xi\delta\upsilon,\label{kkee3}  \\
\delta\xi &=& h,  \label{kkee4}  \\
\bar{\delta \xi} &=& -\xi\tau, \label{kkee5}  \\
D\xi &=& -i\upsilon\xi \label{kkee6} 
\ea
with the constraint
\be  \label{constr}
h+\bar{h} = \xi\bar{\tau}+ \bar{\xi}\tau.
\ee
$K$ is the quantity defined by \eref{kkk}.
If the integrability conditions for the extended system are satisfied identically, the general solution would involve 4 arbitrary constants, but related by one constraint. Hence a 3-parameter group with space-like generators can be expected at most. A first set of integrability conditions is obtained by applying the D-operator on \eref{kkee1}-\eref{constr} and substitution the first derivatives:    
\ba
\xi DK &=& 0,  \\
\bar{\xi} DK &=& 0, \\
(\bar{\delta}\bar{\xi} + \delta\xi) K + \xi(\delta K - K \bar{\tau}) 
+  \bar{\xi}({\bar\delta} K - K \tau)  &=& 0,  \\
D\bar{h} - i \bar{\xi}\bar{\delta}\upsilon  &=& 0, \\
-\delta\bar{\xi} \tau + \bar{\xi}(\bar{\delta}\bar{\tau}- 2\tau\bar{\tau}) + \delta\bar{h} &=& 0, \\
-\bar{\delta}\bar{\xi} \tau - \xi K - \bar{\xi}\bar{\delta}\tau  
+ \bar{\delta}\bar{h} &=& 0.  
\ea
Since $\xi\bar{\xi} \neq 0$, the first two equations require $DK =0$, thus 
$K$ cannot change in the null direction. Taking the constraint \eref{constr} into account, the third equation can be written
\be \label{kke4}
\xi \delta K + \bar{\xi}\bar{\delta}K =0.
\ee
This is again an algebraic equation for $\xi$, but with already satisfied solvability conditions.

\subsection{$\delta K \neq 0$.}
We assume $\delta K \neq 0$ and treat the case $\delta K = 0$ later.
Equation \eref{kke4} is then solved by
\be \label{kke5}
\xi  = 2iw\bar{\delta} K
\ee
with a real function $w$. Introducing this ansatz in the original system 
\eref{kke1}-\eref{kke3}, one obtains
\be \label{kke6}
\bar{\delta}K Dw =0,
\ee
\be \label{kke7}
\delta w \delta K + w(\delta\delta K + \bar{\tau}\delta K) =0,
\ee
\be \label{kke8} \delta w \bar{\delta}K - \delta K \bar{\delta}w =0.
\ee
The first equation requires $Dw = 0$, i.e. $w$ remains unchanged when progressing in the null direction. The change of $w$ in spatial directions is determined by the second equation, while the third one implies  $Re(\delta w)Im(\delta K) = 0$ and either restricts $\delta w$ or $\delta K$. 
We study these integrability conditions by writing them in coordinate form.  
Since $h_{ik}=0$, it follows from \eref{eq1} that the $\gamma_{AB}$  depend only on $x_2,x_3$. The Killing equations \eref{eq0} read in coordinate form:
\be \xi^B_{,1}=0, \ee
\be
\gamma_{AB,C}\xi^C + \gamma_{BC}\xi^C_{,A} +\gamma_{AC}\xi^C_{,B}= 0.
\ee
With help of coordinate transformations \eref{tr1b}, $\xi^B= \delta^B_3$ can be reached. The metric then satisfies $\gamma_{AB,3}=0$.
The transformations that are still allowed form a subgroup  of \eref{tr1a},\eref{tr1b}:
\be \label{tr2}
x'_1 = f(x_1,x_2), ~x'_2 = g(x_2), ~ x'_3 = x_3 + h(x_2)
\ee
with $f_{,1}g_{,2} \neq 0$. Using \eref{tr2}, we may obtain
\be \label{me00x} ds^2 = dx_2^2 + m(x_2)dx_3^2
\ee
with a single positive function $m(x_2)$.   
The spatial components $t^2,t^3$ of the triad may be chosen as $t^2=\frac{i}{\sqrt{2}},~t^3=\frac{1}{\sqrt{2m}}$. From \eref{0ka} and $\xi^k =\delta^k_3$ then follows 
\be
\xi = \sqrt{m/2}.
\ee
Comparison with \eref{kke5} yields
\be
w = \frac{m}{2K_{,2}}.            
\ee
$K$ calculated for the metric \eref{me00x} can be written
\be
K = (-2mm_{,22} +m_{,2}^2)/m^2 
\ee
and depends on $x_2$ only. 
Since $|\delta K| \neq 0$ was assumed, we have $K_{,2} \neq 0$.

Discussing the integrability conditions \eref{kke6}-\eref{kke8} becomes now trivial. 
\eref{kke6} is obviously satisfied. Since $Re~\delta w =0$, \eref{kke8} holds.
A short calculation shows that also \eref{kke7} is satified.  
We have a $G_1$ as isometry group with a spacelike generator, together with a $G_\infty$:
\be  \label{hor2}
\bold{G_\infty x~G_1I}:~~~E=1,~F=0,~ G=m(x_2).
\ee
\subsection{$\delta K = 0$}
In the case $\delta K =0$ the equation \eref{kke4} is trivially solved. Together with $DK=0$, the invariant $K$ is then a \emph{constant} on the null surface. The remaining three integrability conditions are also satisfied, if the initial extended system is considered.
Hence, as in the case $\delta K \neq 0$, also for horizons with $K = const$ all integrability conditions of the extended system are satisfied identically. From  Eisenhart's theorem one expects at most a $G_3$ group of spatial motions, together with the $G_\infty$. 

All these facts are well-known from the differential geometry of two-dimensional surfaces (cf. \cite{My}). With the coordinate transformations \eref{tr1b} we
may obtain normal (or canonical) forms for the metric $(a=const)$, we list also the corresponding KVs:
 
\underline{$K=0$}
\be \label{hor3a}
\bold{G_\infty x~G_3VII_0}: E=a^2,~F=0,~G=a^2 
\ee
\be 
\xxi{0}^i = (f(x_1),0,0),~\xxi{1}^i=(0,0,1),~\xxi{2}^i=(0,1,0),~\xxi{3}=(0,x_3,-x_2).
\ee

\underline{$K=-1/a^2$}
\be \label{hor3b}
\bold{G_\infty x~G_3VIII}: E=a^2/x_2^2,~F=0,~G=a^2/x_2^2    
\ee
\be 
\xxi{0}^i = (f(x_1),0,0),~\xxi{1}^i=(0,0,1),~\xxi{2}^i=(0,x_2,x_3),
\ee
\bea
\xxi{3}^i=(0,2x_2x_3,x_3^2-x_2^2). \nn
\eea

\underline{$K=1/a^2$}
\be \label{hor3c}
\bold{G_\infty x~G_3IX}: E=a^2,~F=0,~G=a^2\cos{x_2}^2
\ee
\be 
\xxi{0}^i = (f(x_1),0,0),~\xxi{1}^i=(0,0,1),~\xxi{2}^i=(0,\cos{x_3},\sin{x_3}\tan{x_2}),    
\ee
\bea
\xxi{3}^i =(0, -\sin{x_3},\cos{x_3}\tan{x_2}). \nn
\eea
The $G_3$ part of the motion group is classified according to Bianchi, 
$G_3VII_0$ is the Bianchi type $G_3VII$ with $q=0$, cf. \eref{bianchi}.     
A horizon metric does not always appear in the canonical form given above. Frequently, the metric 
\be \label{hord}
E=a, ~F=ce^{-x_2},~G=be^{-2x_2}
\ee
($a,b,c =const$) arises as limiting case of space-like $G_2$, $G_3$ and $G_4$ symmetries. Actually, \eref{hord} is a horizon: Using \eref{tr2} with
\be
x'_2 = e^{x_2}\sqrt{ab-c^2}/b,~ x'_3 =x_3 +ce^{x_2}/b
\ee
we obtain the normal form of a $G_\infty \times G_3VIII$ metric with
\be
E' = a'^2/x_2'^2 = (ab-c^2)/(bx_2').
\ee


\subsection{Example: horizons of the Kerr metric}
An example of horizons of the type $G_\infty\times G_1$ is provided by the Kerr metric, which in Finkelstein-Kruskal coordinates can be written
\bea \fl \label{kerr}
ds^2= -(1-\frac{2mr}{\Sigma})dt^2 +2dtdr
-\frac{4amr sin^2\theta}{\Sigma}drd\phi - 2a sin^2\theta drd\phi  \\ \nn
+ \Sigma d\theta^2 +(\frac{(r^2+a^2)^2       - a^2\sin^2\theta \Delta}{\Sigma})
\sin^2\theta d\phi
\eea
with
\be
\Sigma = r^2+a^2 \cos^2\theta, ~ \Delta = r^2-2mr +a^2.
\ee
$\Delta$ has two zeros $r_{-}$ and $r_{+}$. The equation $r=r_{\pm}$ represents
the two horizons.  From \eref{kerr} one obtains the induced metric of the horizons:
\bea 
\gamma_{11} = \frac{a^2\sin^2\theta}{\Sigma_{\pm}},~ \gamma_{12}=0,~ 
      \gamma_{13} = -\frac{2amr_{\pm}}{\Sigma_{\pm}}\sin^2{\theta}, \\ \nn
\gamma_{22} = \Sigma_{\pm},~~~~~~~~ \gamma_{23} =0,~
\gamma_{33} = \frac{4m^2r_{\pm}^2}{\Sigma_{\pm}}\sin^2{\theta}.	      
\eea
A direct calculation gives indeed $det|\gamma_{ik}| = 0.$ A suitable triad in non-adapted coordinates can be found as
\bea
\epsilon^{i} = (2am,0,2m-r_{\pm}), \\
t^i = (0, \frac{i}{\sqrt{2\Sigma_{\pm}}}, \frac{\sqrt{\Sigma_{\pm}}}{2\sqrt{2}mr_{\pm}\sin{\theta}}), \\
\gamma_i = (\frac{1}{2am},0,0), \\
t_i = (-\frac{a \sin{\theta}}{\sqrt{2\Sigma_{\pm}}}, i\sqrt{\frac{\Sigma_{\pm}}{2}},
\sqrt{\frac{2}{\Sigma_{\pm}}}mr_{\pm}\sin{\theta} ).
\eea
The transformation
\be
x'_1 = x_1,~x'_2 = x_2, ~ x'_3 = x_1/(2am)- x_3/(2m-r_{\pm})
\ee
($x_2 \equiv \theta,~x_3 \equiv \phi$) leads to adapted coordinates 
with $\gamma'_{11} =0,\gamma'_{1A} =0$ and to 
\bea
\gamma'_{22}= \Sigma_{\pm}, \\ 
\gamma'_{23} = 0,~  \\
\gamma'_{33} = \frac{4m^2a^4\sin^2{\theta}}{\Sigma_{\pm}}.
\eea
In adapted coordinates the calculation of $K$ is easy, on obtains 
\be
K = \frac{(a^2+r_{\pm}^2)(r_{\pm}^2- 3a^2\cos^2{\theta})}
	 {(r_{\pm}^2+a^2\cos^2{\theta})^3},
\ee
a formula first derived by Smarr \cite{smarr}.

\section{Groups $G_1,G_2$ with spacelike generators}

We now study groups of motions $G_r$ with $r$ \emph{space-like} generators 
$X=\xxi{k}^i\partial_i, k=1...r.$ For the commutators we use Bianchi's notation
\be
[X_i,X_k] =
(\stackrel[i]{}{\xi^l} \stackrel[k]{}{\xi^m_{,l}}- 
\stackrel[k]{}{\xi^l} \stackrel[i]{}{\xi^m_{,l}} )\partial_m
\ee
The maximal number of multiple KS's in a n-dimensional manifold is $n(n+1)/2$,
therefore 6 for the $\mathcal{N}_3.$ 
A $G_6$ is not possible, unlike to a nondegenerate $V_3$. Fubinis theorem 
\cite{fubini} also excludes a $G_5$. 
The reason for the reduced mobility for a $\mathcal{N}_3$ is the existence of the null direction. It causes an inherent asymmetry in null hypersurfaces.

In this section groups $G_1,G_2$ are considered. 
We use adapted coordinates with the line element \eref{gl1}.
This shape of the metric is preserved under the transformations \eref{tr1a},\eref{tr1b}. The Killing equation \eref{keq} can be written  in these coordinates
\be \label{gl2} 
\xi_{,1}^B =0,
\ee
\be  \label{gl3}
\xi^1 \gamma_{AB,1} + \gamma_{AB,C}\xi^C+ \gamma_{AC}\xi^C_{,B} + \gamma_{BC}\xi^C_{,A} = 0.
\ee
Coordinates may further be chosen such that $\xi^i=\delta^i_3$, say.
The metric is then restricted by $\gamma_{AB,3}=0$.
The coordinate transformations preserving this condition as well as the chosen normal form for $\xi^i$ are given by \eref{tr2}.
If only one space-like Killing vector exists, we have a motion $G_1$ with the metric 
\be \label{me00}
\bold{G_1:}~ E=E(x_1,x_2),~ F=F(x_1,x_2),~ G=G(x_1,x_2).  
\ee

\subsection{Abelian groups $G_2I$ with space-like generators }\label{g21}
Taking $\stackrel[1]{}{\xi^i} = \delta^i_3$ as the first Killing field of an Abelian $G_2$, the condition $[X_1,X_2]=0$ or explicitly
\be
\stackrel[1]{}{\xi^i} \stackrel[2]{}{\xi^k_{,i}}- 
\stackrel[2]{}{\xi^i} \stackrel[1]{}{\xi^k_{,i}} =0
\ee
requires $\xxi{2}^k = \xxi{2}^k(x_1,x_2)$. The KE \eref{gl2},\eref{gl3} admit to write $\xxi{2}^1 =\alpha(x_1,x_2),~\xxi{2}^2 = \beta(x_2),~\xxi{2}^3 =\gamma(x_2)$, and lead to 
\bea \label{ke1}
\alpha E_{,1} + \beta E_{,2} + 2E\beta_{,2} +2F\gamma_{,2} &=& 0,\nn \\ 
\alpha F_{,1} + \beta F_{,2} + F\beta_{,2} + G\gamma_{,2} &=& 0, \\
\alpha G_{,1} + \beta G_{,2} &=& 0.\nn
\eea
The functions $\alpha, \beta, \gamma$ transform under \eref{tr2} as
\bea \label{tr3}
\alpha' &=& \alpha f_{,1} + \beta f_{,2}, \nn \\ 
\beta' &=& \beta g_{,2}, \\
\gamma' &=& \gamma + \beta h_{,2}.\nn
\eea
Two cases must be distinguished. Suppose first $\beta \neq 0$ (the case $\beta=0$ is treated in the following subsection). Then, by means of \eref{tr3}, $\beta=1, \gamma=0, \alpha =0$ can be reached, that is, $\xxi{2}^i =\delta_2^i$. 
The residual transformations 
\be \label{tr3a}
x'_1= f(x_1),~ x'_2 = x_2 + const,~ x'_3 = x_3 +const 
\ee
with $f_{,1} \neq 0$ leave $\alpha, \beta$ and $\gamma$ unchanged. 
Integrating the resulting differential equation \eref{ke1} gives a first normal form for a metric admitting a $G_2$ with space-like trajectories: 
\be \label{me0}
\bold {G_2I_{.1}}: ~~~E= a(x_1),~ F = c(x_1),~ G = b(x_1).
\ee
The rotation coefficients of this metric are
\bea \fl
~\rho = -d_{,1}/(4d), \\
\fl ~\sigma = (a_{,1}b^2-b_{,1}ab +2b_{,1}c^2 -2c_{,1}bc)/(4bd)
+i(c_{,1}b-b_{,1}c)/(2b\sqrt{d}),\\
\fl 
~\nu = (b_{,1}c-c_{,1}b)/(2b\sqrt{d}), ~ \tau = 0.
\eea
with $d= ab-c^2$. The differential invariants can be written if 
$|\sigma|= \sqrt{p}/(4d) \neq 0$:
\bea  
 j &=& -  d_{,1}/\sqrt p, \\
 I_1 &=& - 8(d/p)^{3/2}(a_{,11}[c_{,1}b-cb_{,1}] +c_{,11}[b_{,1}a-ba_{,1}] \nn \\
 && + b_{,11}[a_{,1}c-ac_{,1}]), \\
 I_2 &=& j_{,1}/\rho,\\
 J &=& 0
\eea
with  
\be \label{ppp}
p= (a_{,1}b-ab_{,1})^2 + 4(a_{,1}c-ac_{,1})(b_{,1}c-bc_{,1}).
\ee 
Two special cases must be considered. In the case $|\sigma| = \sqrt{p}/(4d)=0$ (or $p=0$) and  $\rho \neq 0$ (or $d_{,1} \neq 0$), the same metric applies (now denoted $\bold{G_2I_{.2}}$), but no invariants exist. If additionally $d=const$, this results in a \emph{constant} metric with an additional space-like KV, a case treated later. 
Residual coordinate transformations are 
\eref{tr3a}. 

\subsection{Abelian groups $G_2I$ with spacelike generators coplanar with the null direction }\label{g22}
If $\beta =0$, the situation is slightly more complicated. We suppose that $\alpha \neq 0$, otherwise $\xxi{2}^i$ would be equivalent to $\xxi{1}^i$ and no $G_2$ exists. Then, with $\alpha' = \alpha f_{,1}$, we reach $\alpha'=1$, and the simplified equation \eref{ke1} can be integrated with the result:
\bea \label{me1} 
E &=& E_0(x_2) -2\gamma_{,2} x_1 F_0(x_2) + \gamma_{,2}^2 x_1^2 G_0(x_2), \nn \\
F &=& F_0(x_2) - \gamma_{,2} x_1 G_0(x_2), \\
G &=& G_0(x_2). \nn
\eea
Since $\xxi{2}^i$ must be spacelike, we have $\gamma(x_2) \neq 0$.
The still available coordinate freedom is 
\bea \label{tr4}
x'_1 &=& x_1 + f(x_2),\nn \\
x'_2 &=& g(x_2),\\
x'_3 &=& x_3 + h(x_2) \nn
\eea
with three free functions $f,g,h$ of $x_2$. A little calculation shows that under \eref{tr4} $E_0,F_0,G_0$ change as   
\bea
g_{,2}^2 E'_0 &=&  E_0 +2F_0(f\gamma_{,2}- h_{,2}) +G_0 (f \gamma_{,2} -h_{,2})^2 , \nn \\
g_{,2} F'_0 &=& F_0 + G_0(f \gamma_{,2} - h_{,2}), \\
G'_0 &=& G_0  \nn
\eea
Also, transforming the Killing vector $\xxi{2}^i =(1,0,\gamma(x_2))$ leads to $\gamma'(x_2') = \gamma(x_2)$. We can use these transformations to obtain a certain normal form of the metric. e.g., $E_0' =1,~F'_0=0$. Assuming \underline{$\gamma_{,2} \neq 0$}, we choose
\be
f = \frac{1}{\gamma_{,2}}(h_{,2} - F_0/G_0), ~ g_{,2}^2 = (E_0G_0- F_0^2)/G_0  
\ee
to obtain 
\bea \label{ke2}
 \bold {G_2I_{.3}}:\\
E &=& 1 + x_1^2\gamma_{,2}^2m(x_2), \nn \\
F &=& x_1\gamma_{,2}m(x_2),  \\
G &=& m(x_2). \nn
\eea
The still available coordinate freedom
\be \label{tr5}
x'_1 = x_1 -h_{,2}/\gamma_{,2},~ x'_2 =x_2 +const,~x'_3 =x_3 + h(x_2) 
\ee
does not change this metric. The rotation coefficients are
\bea
\rho = 0,~ |\sigma| = \frac{\sqrt{m}\gamma_{,2}}{2}, \\
\sigma =i\frac{\sqrt{m}\gamma_{,2}}{2}, ~\nu = -\frac{\sqrt{m}\gamma_{,2}}{2},~
\tau =i\frac{m_{,2}}{2\sqrt{2}m}  
\eea
For the invariants we obtain
\bea
j=0,~ I_1 = -2 \gamma_{,2}/|\gamma_{,2}|= -2sgn(\gamma_2),\\
N= (cos(s)-sin(s))m_{,2}/(\sqrt{2}m).
\eea
The surfaces $x_1= const$ have the Gaussian curvature
\be
K = \frac{-2mm_{,22}+ m_{,2}^2}{4m^2}.
\ee
It is noteworthy that the metric \eref{ke2} allows a \emph{continous number of spacelike Killing fields} commuting with the first two:
\be
\xxi{3}^i = (C_{,2}/\gamma_{,2},0,C),
\ee
where $C$ is an arbitrary function of $x_2$. $\xxi{2}^i$ is included for $C=\gamma$.
\noindent

We treat now the case \underline{$\gamma = const$}. Here the transformation functions $h,g$ satisfying
\be
h_{,2} = F_0/G_0, ~ g_{,2}= \sqrt{(E_0G_0-F_0^2)/G_0}
\ee
result in
\be \label{me2}
~E=1,~ F=0,~ G=m(x_2).
\ee
The residual transformations 
\be
x'_1 = x_1 + f(x_2),~x'_2= x_2,~x'_3= x_3 + const
\ee
leave \eref{me2} invariant. \eref{me2} is the metric of the horizon \eref{hor2}.

The two types of Abelian groups $G_2I$ with space-like generators discussed here and in the previous subsection can be characterized by the quotient
\be
Q = \frac{\xxi{1}^i\xxi{2}^k\gamma_{ik}}{\sqrt{\xxi{1}^i\xxi{1}^k\gamma_{ik}} 
\sqrt{\xxi{2}^i\xxi{2}^k} \gamma_{ik}}.
\ee
In the first case ($\beta \neq 0$)  we have in adapted coordinates
\be
Q = \gamma_{23}/\sqrt{|\gamma_{AB}|+ \gamma^2_{23}} < 1,
\ee
in the second case ($\beta = 0$)
\be
Q = 1.
\ee
$Q$ is the cosinus of the angle between the space-like directions $\xxi{1}^i$ and $\xxi{2}^i$, and $Q=1$ tells us that this angle is zero.
Nevertheless, the two KVs represent different directions, they form a so-called 
\emph {null angle} with each other\footnote {The concept of null angles was introduced by R. Penrose in his Les Houches lectures 1973 \cite{pen1}.}.
The geometrical interpretation is that the two space-like directions span a light-like surface element, i.e. that the null direction $\epsilon^i$ can be represented as a linear combination of the KVs. In this case the vectors are called \emph{coplanar} with the null direction. The criterion for two vectors $a^i,b^i$ to be coplanar with the null direction $\epsilon^i$ is
\be
\epsilon_{ikl}a^ib^k\epsilon^l = 0,
\ee
where $\epsilon_{ikl}$ is the totally antisymmetric Levi-Civita symbol with $\epsilon_{123}=1$.

\subsection{Non-Abelian groups $G_2II$ with spacelike generators}\label{g23}
Non-Abelian groups of order 2 are denoted by $G_2II$ and satify the commutator relation 
 \be
[X_1,X_2]=X_1.
\ee
If the first KV is transformed into $\xxi{1}^i=\delta_3^i$, this relation gives  
\be
\xxi{2}^i = (\alpha(x_1,x_2),\beta(x_2), x_3+ \chi(x_2))
\ee
with a new function $\chi(x_2)$. The transformations \eref{tr2} change $\alpha, \beta,\chi$:
\be
\alpha'= f_{,1}\alpha +f_{,2}\beta,~ \beta'=g_{,2}\beta,~\chi'=\chi -h +h_{,2}\beta.
\ee
Again two cases must be distinguished. Assuming first $\beta \neq 0$, we may reach $\alpha'=0,\beta'=1,\chi'=0$. Integrating the KE for $\xxi{2}^i$ yields the metric
\be \label{ke3}
\bold {G_2II_{.1} :}~E= a(x_1), ~F = c(x_1)e^{-x_2},~ G = b(x_1)e^{-2x_2}.
\ee
Still allowed are the coordinate transformations
\be \label{tr6}
x_1' = f(x_1), ~ x_2' = x_2 + g_0, ~ x_3' = x_3 + h_0 e^{x_2},
\ee
where $f_{,1} \neq 0$ and an inverse function $x_1= \varphi(x_1')$, $g_0$ and $h_0$
are constants. By means of \eref{tr6} the metric \eref{ke3} is transformed into
\be \fl \label{ke4}
E' = a'(x_1')-2c'(x_1')h_0 + b'(x_1')h_0^2, 
~ F' = e^{g_0 -x_2'}(c'(x_1')-h_0b'(x_1')), ~ G' = b'(x_1') e^{2(g_0- x_2')}
\ee
where $a'(x_1') \equiv a(\varphi(x_1'))$ etc.

The rotation coefficients for the metric \eref{ke3} are
\bea 
\rho = -d_{,1}/(4d), \\
\sigma =(a_{,1}b^2-b_{,1}ab+2b_{,1}c^2-2c_{,1}bc)/(4bd)+i(bc_{,1}-cb_{,1})/(2b\sqrt{d}), \\
\nu = (b_{,1}c-c_{,1}b)/(2b\sqrt{d}),\\
\tau = -i\sqrt{b/(2d)}, ~d \equiv ab-c^2.
\eea
The invariants can be written, if $d_{,1} \neq 0$:
\bea \fl
j = -d_{,1}/\sqrt{p}, \\ 
\fl I_1 = 8(d/p)^{3/2}(a_{,11}(b_{,1}c-c_{,1}b)+ b_{,11}(c_{,1}a-a_{,1}c)+ c_{,11}(a_{,1}b-b_{,1}a)), \\
\fl I_2 = 2\frac{d}{\sqrt{p}}(2\frac{d_{,11}}{d_{,1}} - \frac{p_{,1}}{p}),   
\eea
The function $p(x_1)$ is again given by \eref{ppp}.
The invariants also include $J$, which, unlike $I$, is \emph{nonlinear in the second derivatives of the metric}. $J$ describes the change of the geometry in the transverse directions and has a relatively complicated structure: 
\be \label{ke4a}
J = -128 e^{is/2}\frac{UV}{\sqrt{2b}p^3d_{,1}^2} 
\ee 
with $e^{is/2}= \sqrt{\sigma/|\sigma|}$. $U= U_1+iU_2$ is complex with 
\bea
\fl  U_1 = 2bd^2d_{,1}(a_{,11}[bc_{1}-cb_{,1}] +b_{,11}[ca_{,1}-ac_{,1}]    
    +c_{,11}[ab_{,1}-ba_{,1}]) +dd_{1}p(ab_{,1}-ba_{,1}), \nn  \\
\fl U_2 = 2bd^{5/2}(la_{,11} + mb_{,11} +2 n c_{,11})  \nn   \\
\fl~~~~~~+pd^{1/2}(-b^3a_{,1}^2+b(c^2-3ab)a_{,1}b_{,1} + 4b^2ca_{,1}c_{,1} \nn  \\
\fl~~~~~~-ac^2b_{,1}^2 +2c(c^2+ab)b_{,1}c_{,1} +2b(ab-3c^2)c_{,1}^2). \nn
\eea  
We have used the abbreviations:
\bea
l= ba_{,1}b_{,1} -ab_{,1}^2 + 2cb_{,1}c_{,1} -2bc_{,1}^2 \nn \\
m= aa_{,1}b_{,1} -ba_{,1}^2 + 2ca_{,1}c_{,1}- 2ac_{,1}^2 \nn \\
n = -2ca_{,1}b_{,1} + ab_{,1}c_{,1} +ba_{,1}c_{,1}.\nn
\eea
Finally, $V$ is a real function:
\be
V = d(la_{,11}+mb_{,,1}+ nc_{,11}) -p(a_{,1}b_{,1}-c_{,1}^2) \nn
\ee

Again, two special cases must be considered.

In the case $d_{,1}=0$, the metric \eref{ke3} admits - in addition to $j=0$ - only the invariant $I_1$:
\be
I_1 = - sgn(c)\frac{2B}{\sqrt{d}A^{3/2}}
\ee
with
\bea
A=(a_{,1}b+b_{,1}a)^2 -4a_{,1}b_{,1}c^2  \nn \\
B=4dc^2(a_{,11}b_{,1}-b_{,,11}a_{,1}) \nn
+ ba_{,1}^2(a_{,1}b^2+b_{,1}ab-4b_{,1}c^2) \nn \\
- ab_{,1}^2(b_{,1}a^2+a_{,1}ab-4a_{,1}c^2). \nn
\eea
The surfaces $x_1=const$ 
have the constant negative curvature $K= -b/d$.

The case $|\sigma|=0$ implies $p=0$ and leads to
\be
a(x_1) = k_1b(x_1),~c(x_1)= k_2b(x_1)
\ee
with the constants $k_1,k_2,~k_1>k_2^2$. We first assume $b_{,1} \neq 0.$ The metric is
\be \label{ke3a}
\bold {G_2II_{.3} :}~E= k_1b(x_1), ~F = k_2b(x_1)e^{-x_2},~ G = b(x_1)e^{-2x_2}.
\ee
Rotation constants are 
\be
\rho=-b_{,1}/(2b),~ \sigma=\nu =0,~ \tau= -i/(\sqrt{2b}\sqrt{k_1-k_2^2}).
\ee
There exist no invariants. The surfaces $x_1=const$ have the negative Gaussian curvature $K= -1/(b(k_1-k_2^2))$. 

If however $b_{,1} = 0$, the Gaussian curvature $K$ is an invariant and the metric \eref{ke3a} degenerates to a horizon. \eref{ke3a} is then equivalent to \eref{hor3b}. 

\subsection{Non-Abelian groups $G_2II$ with spacelike generators coplanar with the null direction}\label{g24}

Finally, the case $\beta=0$ has to be treated. Here the changes 
\be 
\alpha' = f_1 \alpha, ~ \chi' = \chi - h(x_2)
\ee
are possible, leading to (since obviously
$\alpha \neq 0$) $\alpha'=1, \chi' = 0$. The KE can be integrated with 
the result 
\be \label{ke5} 
E= E_0(x_2), ~ F = F_0(x_2)e^{-x_1}, ~ G = G_0(x_2)e^{-2x_1}.
\ee
The coordinate transformations 
\be \label{tr7}
x_1' = x_1 + f(x_2),~ x_2' = g(x_2),~ x_3' = x_3,
\ee
leave this metric structure invariant,
but the functions $E_0, F_0, G_0$ change as 
\be
E'_0 = E_0/g_2^2, ~ F'_0 = F_0 e^{f}/g_2, ~ G'_0 = G_0 e^{2f}.
\ee
With suitable $f(x_2), g(x_2)$ we can obtain a certain normal form ($k > 1$): 
\be \label{ke6}
\bold{G_3VIII._4}: E = 1,~ F= e^{-x_1},~ G = k(x_2)e^{-2x_1}.
\ee
An integration constant was absorbed in a changed $x_2$. 
The residual transformations leaving this normal form invariant are
\be
x'_1 = x_1,~  x'_2 = x_2 + const, ~ x'_3 = x_3.
\ee
The rotation coefficients for \eref{ke6} are
\bea
\rho= \frac{1}{2},~ |\sigma|= \frac{1}{2}\frac{\sqrt{k}}{\sqrt{k-1}},~
\sigma = \frac{1}{2} + \frac{i}{2\sqrt{k-1}}, \\
\nu = -\frac{1}{2\sqrt{k-1}}, 
~ \tau =  \frac{ik_{,2}}{2^{3/2}\sqrt{k(k-1)}}.
\eea
The invariants are
\bea
j= \sqrt{1-1/k},~I_1 = -2/\sqrt{k},~ I_2 = I_3 = J =0,\\
L=\frac{k_{,2}e^{is/2}}{(k-1)\sqrt{2k}},
~M=-i\frac{k_{,2}e^{is/2}}{(k-1)^{3/2}\sqrt{2k}},
\eea 
note $e^{is/2}= (\sigma/|\sigma|)^{1/2}$.
The surfaces $x_1=const$ have the Gaussian curvature
\be
K = -\frac{k_{,22}}{2(k-1)}  + \frac{(k_{,2})^2}{4(k-1)^2}.
\ee
We may ask if the metric \eref{ke6} - as its notation suggests - admits an 
additional KV, say $\xxi{3}^i =(A,B,C)$. The KE for this vector represents
a system of differential equations for $A,B,C$, solved by          
\be
\xxi{3}^i = (C_{,3},0,C),
\ee 
$C$ depends only on $x_3$. The commutator relations are given by
\be \fl
|X_1,X_2|= X_1,~ |X_1,X_3|= (C_{,33},0,C_{,3}),~ |X_2,X_3|=(x_3C_{,33},0,x_3C_{,3}-C).  
\ee
If $C$ is a constant, $\xxi{3}^i$ essentially reduces to $\xxi{1}^i$.
The choice $C=x_3$ reproduces $\xxi{2}^i$, but $C= x_3^2$ or $\xxi{3}^i =
(2x_3,0,x_3^2)$ leads to the commutator relations
\be
|X_1,X_2| = X_1,~ |X_1,X_3| = 2X_2,~ |X_2,X_3| = X_3,
\ee
corresponding to a three-parameter Lie group of the Bianchi type VIII, with operators $X_i$ as given by Petrov \cite{petrov}.

\section{Groups $G_3$ with spacelike generators}

\subsection{Overview of Groups $G_3$ }
The classification of three-parameter Lie groups was given by Bianchi \cite{bianchi}, see also Petrov \cite{petrov}:
\bea \label{bianchi}
I.~ [X_1,X_2] &=& 0,~ [X_1,X_3] = 0,~ [X_2,X_3] =0 \nn \\
II.~ [X_1,X_2] &=& 0,~ [X_1,X_3] = 0,~ [X_2,X_3] =X_1 \nn \\
III.~ [X_1,X_2] &=& 0,~ [X_1,X_3] = X_1,~ [X_2,X_3] =0 \nn \\
IV.~ [X_1,X_2] &=& 0,~ [X_1,X_3] = X_1,~ [X_2,X_3] = X_1+X_2 \nn \\
V.~ [X_1,X_2] &=& 0,~ [X_1,X_3] = X_1,~ [X_2,X_3] =X_2  \\
VI.~ [X_1,X_2] &=& 0,~ [X_1,X_3] = X_1,~ [X_2,X_3] = qX_2~ (q \neq 0,1) \nn \\
VII.~ [X_1,X_2] &=& 0,~ [X_1,X_3] = X_2,~ [X_2,X_3] = -X_1+qX_2~ (q^2<4) \nn \\ 
VIII. ~ [X_1,X_2] &=& X_1,~ [X_1,X_3] = 2X_2,~ [X_2,X_3] = X_3 \nn \\ 
IX.~ [X_1,X_2] &=& X_3,~ [X_1,X_3] = -X_2,~ [X_2,X_3] =X_1  \nn     
\eea
The groups $I$ through $VII$ admit an Abelian $G_2$ as subgroup. We can use the results of sections \ref{g21} and \ref{g22}. 
Thus we assume $\xxi{1}^i = \delta^i_3$ and $\xxi{2}^i = \delta^i_2$ as the first two commuting Killing fields not coplanar with the null direction (denoted as first cases).
Then the resulting metric components $E,F,G$ depend only on a single coordinate $x_1$.

We have also to ask for every Bianchi class, if KVs coplanar with the null direction do exist (denoted as second cases). We then follow section \ref{g22} and write the first two KVs as $\xxi{1}^i = \delta^i_3$ and $\xxi{2}^i = (1,0,\gamma(x_2))$.

The third Killing field is always written $\xxi{3}^i = (A,B,C)$.      
The calculations will also show whether there are any null vectors among the Killing vectors, i.e. whether the corresponding metrics represent horizons. 

\subsection{$G_3,$ Bianchi type I (first case)} 

Assume $\xxi{1}^i = \delta^i_3$ and $\xxi{2}^i = \delta^i_2$.
The commutator relations $|X_1,X_3|=0,~|X_2,X_3|=0$ 
state that the components $A,B,C$ of the third KV can depend at most on $x_1$. 
The  first part of the KE \eref{gl2} for the third KV shows that $B$ and $C$ are constants. From the second part \eref{gl3} one obtains
\be
A E_{,1} =0,~ A F_{,1} =0,~ A G_{,1} =0
\ee
i.e. a constant metric in case $A \neq 0$. $A=0$
is ruled out, since the third KV would be a linear combination of the first two:
$\xxi{3}^i = B\xxi{2}^i + C\xxi{1}^i$, and no $G_3$ exists. With the residual transformation \eref{tr3a} we obtain $A=1$. With $X'_3 =X_3-BX_2-CX_1$ we can also remove the components $B$ and $C$ from the third KV without violation of the commutator relations, thus $\xxi{3}^i = \delta^i_1$, which is null. We have the horizon \eref{hor3a} with a vanishing invariant $K$.

\subsection{$G_3,$ Bianchi type I (second case)}
Are KVs possible for the Bianchi type I, which are coplanar with the null direction? Taking $\xxi{1}^i=\delta^i_3, ~\xxi{2}^i=(1,0,\gamma(x_2))$,
a third Killing field 
commuting with the first two requires that the components $A,B,C$ depend only on $x_2$ and that $B\gamma_{,2}=0$. The latter condition shows that three cases must be considered: 
\bea
\gamma_{,2} \neq 0, B=0, \nn \\ \gamma_{2}=0,B \neq 0, \nn \\ \gamma_2=B=0.\nn  
\eea
Assuming first \underline{$\gamma_{,2} \neq 0$, $B=0$}. $A \neq 0$ is required, since otherwise $\xxi{3}^i = C \xxi{1}^i$ and only a $G_2$ exists. With $\xxi{3}^i= (A(x_2),0,C(x_2))$ we write the KE
\be \nn
AE_{,1} +2F C_{,2}=0,~AF_{,1}+GC_{,2} =0, ~ AG_{, 1} =0.     
\ee 
Using the representation \eref{me1} of the metric, we find the further restriction $-A\gamma_{,2} + C_{,2} = 0.$ Since $\gamma_{,2} \neq 0$, we have
$ \xxi{3}^i = (C_{,2}/\gamma_{,2}, 0, C).$ 
The third KV depends on a - by means of \eref{tr4} - non-removable arbitrary function $C(x_2) \neq \gamma$. 
This case was already treated, see \eref{ke2}. 

The next case is \underline{$\gamma_{,2}=0,B \neq 0$}. Here a look at \eref{me1} shows that the components of the metric depend only on $x_2$. The KEs for $\xxi{3}^i$ are
\be \label{me4}
E_{,2}B + 2EB_{,2} + 2FC_{,2} =0,~ F_{,2}B +FB_{,2} + GC_{,2},~G_{,2}B =0.
\ee
Since $B \neq 0$, we have $G=b=const$. The equations \eref{me4} 
can be further integrated. With constants $a,b,c$ and functions $l(x_2), m(x_2)$ we have
\ben E = l^2(a+bm^2-2cm),~ F = l(c-bm),~ G = b.     \een
However, the functions $l,m$ are spurious and can be removed by a coordinate transformation \eref{tr4}: $ f_{,2} = -l(x_2)A(x_2),~ g_{,2} = l(x_2),~ h_{,2} = -m(x_2)l(x_2)$. The final result is $E=a,~F=c,~G=b$, that is a constant metric, and $\xxi{3}^i= (0,1,0)$, corresponding to the horizon metric \eref{hor3a}. 

Finally we have the case \underline{$\gamma_{,2}=0,B=0$}. Again the metric depends only on $x_2$:
\be \label{me6} 
E  =  k(x_2),~ F = l(x_2),~G= m(x_2).
\ee 
The third KV is a combination of the first 
\be
\xxi{3}^i = A \xxi{2}^i + (C-\gamma)\xxi{1}^i, \nn
\ee
where $C$ (like $\gamma$) is a constant from the KE for $\xxi{3}^i$, thus no $G_3$ with spacelike generators exists here.
The metric admits however an isotropic KV, connected to $\xxi{1}^i$ 
and $\xxi{2}^i$ via $\epsilon^i = \gamma\xxi{1}^i - \xxi{2}^i$.
It is therefore also a horizon, identical to metric \eref{me2} with slightly changed coordinates. 



\subsection{$G_3$, Bianchi type II } 
We first assume that $X_1$ and $X_2$ are not coplanar with the null direction, hence $\xxi{1}^i = \delta^i_3$ and $\xxi{2}^i = \delta^i_2$.
$|X_1,X_3| =0$ says that $A,B,C$ are independent of $x_3$, and $|X_2,X_3|= \delta^i_3$ gives the further restriction $(A,B,C)= (A_0(x_1),b_0,x_2+c_0)$ with constants $b_0,c_0$. $A_0 \neq 0$ is required for a non-singular metric. With the transformation \eref{tr3a} we have $A_0=1,~c_0=0$, also $b_0=0$ can be assumed without loss of generality. Then the KE for $\xxi{3}$ leads to 
\bea E_{,1} + 2F=0, ~F_{,1} = -G,~ G_{,1}  =0, \nn \eea 
and integrated to 
\bea \label{k31}
\bold {G_3II_{.1}:} \nn \\
E &=& a -2cx_1 +bx_1^2, \nn \\
F &=& c -bx_1,  \\
G &=& b. \nn 
\eea
$a,b,c$ are integration constants. With a suitable coordinate change $x_1'= x_1 + const$, $c$ disappears. 
The rotation cofficients are:
\bea
\rho = 0,~ |\sigma| =|b|/(2\sqrt{ab-c^2}), \\ 
\sigma = -ib/(2\sqrt{ab-c^2}), \nu =b/(2\sqrt{ab-c^2}), \tau = 0.
\eea
Invariants are
\be j=0,~ I_1 = 2, N = 0,
\ee
the surfaces $x_1=const$ are flat. 
Notice that $X_1$ and $X_3$ are coplanar with the null direction, which was 
not to be expected from the outset. 
Nevertheless, we have still to deal with the second case that $X_1,X_2$
is coplanar with the null direction. Again $|X_1,X_3| = 0$ implies that $A,B,C$ are independent of $x_3$. From $|X_2X_3|= \delta^i_3$ we have $A_1 =0,~B\gamma_{,2} = -1$, hence $B \neq 0, \gamma_{,2} \neq 0$. 
The transformatiom \eref{tr4} allows $A=0,~B=1,~C=0$, hence $\gamma_{,2}= -1$.
The KE for $\xxi{3}^i$ gives in combination with \eref{me1} 
\ben  E = a +2cx_1 +bx_1^2, ~ F = c +bx_1,~ G = b.    
\een
But this is essentially the same metric as \eref{k31}: The roles of $X_1$ and $X_2$ are reversed in the two cases.

\subsection{$G_3$, Bianchi types III,V,VI (first case)} 
The Bianchi types $III,~V,~VI$ have the commutator relations
\bea
|X_1,X_2| = 0,~~ |X_1,X_3| = X_{1},~~ |X_2,X_3| = qX_{2}
\eea
with a parameter $q$ that distinguishes the different types:

\be
III: q=0, ~V: q= 1,~ VI: q \neq 0,1.
\ee 
In this section we will treat them together.
The condition $|X_1,X_3| = X_1$ provides $A_{,3}=0,~B_{,3}=0,~C_{,3}=1$. The remaining condition $|X_{2},X_{3}| = qX_{2}$ finally leads to
\be
\xxi{3}^i = (A(x_1),~B = qx_2 +c_1,~C= x_3+c_2)
\ee
with two integration constants $c_1, c_2$.
The KE for $\xxi{3}^i$ would lead to $G=0$, if $A=0$, hence $A \neq 0$ and, after a coordinate transformation \eref{tr3a}, $A=1$. \eref{tr3a} also removes the constant $c_2$; $c_1$  can be removed if $q \neq 0$. Considering also \eref{me0}, the equations for the metric are finally:
\be
E_{,1} +2Eq =0, \\
F_{,1} + F(1+q) = 0,\\
G_{,1} + 2 G = 0.
\ee
The integration gives
\be \label{me7}
E =  a e^{-2qx_1},~ F  =  c e^{ -(q+1)x_1},~ G  =  b e^{-2x_1} 
\ee 
with constants $a,b,c,~ab-c^2>0$.
\noindent
We now specialize in the different Bianchi types.
\\
\underline{Bianchi type $III,~ q=0$, first case}

\noindent
Metric:
\bea \label{k33}
\bold {G_3III{.1}:} \nn \\
E = a, ~ F = c e^{-x_1},
~ G = b e^{-2x_1},\\ \label{k34}
|g_{AB}|= (ab-c^2)\exp(-2x_1)
\eea
\noindent
Rotation coefficients:

\be
\rho= \frac{1}{2},~ \nu = -\frac{c}{2\sqrt{ab-c^2}},~\sigma=\frac{1}{2} -i\nu,~\tau =0.             
\ee
\noindent
Invariants:   
\bea  
j= \sqrt{(ab-c^2)/(ab)},\\
I_1= -2c/\sqrt{ab},~I_2=0,~I_3=I_1^2-4(1-j^2)=0,~M=0.  
\eea
The surfaces $x_1=const$ are flat.

\underline{Bianchi type $V,~q=1$, first case} 

\noindent
Metric:
\bea \label{k35} 
\bold {G_3V_{.1}:} \nn \\
E = a e^{-2x_1},~ F = c e^{-2x_1},~ G = b e^{-2x_1}.
\eea
\be \label{k36}
|g_{AB}|= (ab-c^2)\exp(-4x_1).
\ee
\noindent

Rotation coefficients:

\be 
\rho = 1,~\sigma =0,~\nu = 0,~\tau = 0.
\ee
\noindent
No invariants, the surfaces $x_1=const$ are flat.
\\

\underline{Bianchi type $VI,~q \neq 0,1$, first case}

\noindent
The metric is \eref{me7}, denoted $G_3VI_{.1}$.

\noindent
Rotation coefficients:
\bea
\rho = (q+1)/2,~ \sigma =      (1-q)(1 + ic/\sqrt{ab-c^2})/2,\\
\nu = c(q-1)/(2\sqrt{ab-c^2}),~ \tau = 0.
\eea
Invariants:
\be
j = \frac{q+1}{q-1}\sqrt{1-c^2/(ab)},~ I_1 = 2c/\sqrt{ab},~ I_2 = 0,~ J=0
\ee

\subsection{$G_3$, Bianchi types III,V,VI (second cases)} 

If $\xxi{1}^i, \xxi{2}^i$ are coplanar with the null direction, the commutator relations for the third KV lead to
\bea \fl \nn
A=a_0(x_2)+q x_1,~B=b_0(x_2),~C= x_3 + c_0(x_2),~ \gamma(q-1)+b_0(x_2) \gamma_{,2} =0.
\eea
For a regular metric, $b_0(x_2)$ must be nonzero.
This follows from the KE with $\xxi{3}^i$. The coordinate transformations \eref{tr4} allow to obtain $a_0(x_2)=0,~b_0(x_2)=1,~c_0(x_2)=0$. We now write the KE for the simplified third KV:
\bea \fl \nn
qx_1E_{,1} + E_{,2} = 0, ~ 
qx_1F_{,1} + F_{,2} + F = 0, ~ 
qx_1G_{,1} + G_{,2} + 2G = 0
\eea
and compare the result with \eref{me1}. 
Beside 
\be
\gamma = ke^{x_2}, k = const
\ee
we finally obtain
\noindent
\bea  
E &=& a -2kc(1-q)x_1 e^{-qx_2} + k^2 b(1-q)^2x_1^2 e^{-2qx_2}, \nn \\
F &=&  c e^{-x_2}-kb(1-q)x_1 e^{-(1+q)x_2},  \label{k37} \\
G &=& b  e^{-2x_2}  \nn      
\eea
with
\be \label{k38}
EG-F^2= (ab-c^2) e^{-2x_2}.
\ee

We specialize again in the different Bianchi types.

\underline{Bianchi type $III,~ q=0$, second case}
\\

\noindent
Metric:
\bea  \label{k39}          
\bold {G_3III_{.2}:}    \nn    \\ 
E = a-2kcx_1 +bk^2x_1^2,  \\
F = (c-bkx_1)e^{-x_2},  \\
G = b e^{-2x_2}.
\eea

\noindent
Rotation coefficients:
\be \fl
\rho=0,~\nu=bk/(2\sqrt{ab-c^2}),~\sigma= -i\nu,~\tau=i\sqrt{b/(2(ab-c^2))}
\ee

\noindent
Invariants:
\be
j = 0,~I_1 = 2 sgn(k),~N= 4\sqrt{b}(\cos(s/2)-\sin(s/2))/\sqrt{2(ab-c^2)}.
\ee
The surfaces $x_1=const$ have the constant negative Gaussian curvature
\be
K = -b/(ab-c^2).
\ee

For $k=0$ the metric \eref{k39} degenerates into a horizon and can be transformed into \eref{hor3b}.

\underline{Bianchi type $V,~q=1$, second case} 

\noindent 
The metric is given by 
\bea \label{k311}  
E = a,~ F = c e^{-x_2},~ G = b e^{-2x_2}.
\eea

\noindent
Rotation coefficients:
\be 
\rho=0,~\sigma=0,~\nu=0,~ \tau = -i\sqrt{b}/\sqrt{2(ab-c^2)}
\ee

\noindent
Invariants: 
\be
K= -b/(ab-c^2) < 0.
\ee
We have a horizon with the negative constant curvature $K$, transformable into \eref{hor3b}.
  
\underline{Bianchi type $VI,~q \neq 0,1$, second case} 

\noindent
The metric is given by \eref{k37}, denoted $G_3VI_{.2}$.

\noindent
Rotation coefficients:
\bea
\rho=0,~\sigma= -i\nu,~ \nu = bk(1-q)\exp(-qx_2)/(2\sqrt{ab-c^2}),\\
~\tau=-i\sqrt{b/(2(ab-c^2))} 
\eea
Invariants:
\bea
j=0,~ I_1= 2 sgn(k(1-q)),\\
N=-4\sqrt{b}(\cos(s/2)-\sin(s/2))/\sqrt{2(ab-c^2)}.
\eea
The surfaces $x_1=const$ have the constant negative Gaussian curvature $K=-b/(ab-c^2)$.

\subsection{$G_3$, Bianchi type IV (first case)} 
The commutator relations
$|X_1,X_3|= X_1,~
|X_2,X_3|= X_1+X_2$
give
$A_{2}=0,~B_{,3}=0,~C_{,3}=1$
and $A_{,2} = 0,~B_{,2} =1,~C_{2}=1$.
$A=A(x_1)$ must be different from zero since the KE with $\xxi{3}^i$
otherwise leads to a singular metric. With the coordinate  transformation \eref{tr3a} we obtain $A=1$ and remove the constants $c_2,c_3$ in $B=x_2 + c_2,~ C=x_2+x_3 + c_3$. Thus finally we have $\xxi{3}^i = (1,x_2,x_2+x_3)$, along with $\xxi{1}^i=(0,0,1),~\xxi{2}^i=(0,1,0)$. The KE with $\xxi{3}^i$ restricts the metric \eref{me0}:
\bea \label{k312}
\bold {G_3IV_{.1}:}    \nn  \\
E = (a-2cx_1 +bx_1^2) e^{-2x_1}, \\
F = (c-bx_1) e^{-2x_1}, \\
G = b e^{-2x_1}.
\eea
Rotation coefficients:
\be
\rho= 1,~ \sigma= ib/(2\sqrt{ab-c^2}),~ \nu = b/(2\sqrt{ab-c^2}), ~ \tau =0.
\ee
Invariants:
\be
j= 2\sqrt{ab-c^2}/|b|,~I_1=2,~I_2=0,~I_3 = 16(ab-c^2)/b^2, J=0,M=0.
\ee

\subsection{$G_3$, Bianchi type IV (second case)} 
If $\xxi{1}^i,\xxi{2}^i$ are coplanar with the null direction, the commutator relation $|X_1,X_3|= X_1$ returns the same result as in the previous section, 
$A_{2}=0,~B_{,3}=0,~C_{,3}=1$. The second relation $|X_2,X_3|= X_1+X_2$
provides $A_{,1}=1,~B\gamma_{,2}= -1$, so we have
\bea \nn
\xxi{3}^i = (x_1 + a_0(x_2), B(x_2), x_3 +c_0(x_2)).
\eea
Transformations \eref{tr4}  bring $\xxi{3}^i$ into the form
\bea \nn
\xxi{3}^i = (x_1, 1, x_3).
\eea
The KE for this KV yields:
\bea \nn
x_1E_{,1} + E_{,2} = 0, \\
x_1F_{,1} + F_{,2} + F = 0, \nn \\
x_1G_{,1} + G_{,2} + 2G = 0. \nn
\eea
Using the equations \eref{me1} with $\gamma = -x_2 +const$ 
in the KE, one can finally conclude that           
\numparts
\bea \label{k313}
\bold {G_3IV_{.2}:} \nn \\
E = a +2cx_1 e^{-x_2} + bx_1^2 e^{-2x_2}, \\
F = c e^{-x_2} + bx_1 e^{-2x_2}, \\
G = b e^{-2x_2},\\ 
EG-F^2 = (ab-c^2) e^{-2x_2}.
\eea
\endnumparts

\noindent
Rotation coefficients:
\numparts
\bea
\rho=0,~ \sigma=-i\nu,~ \nu = -b\exp(-x_2)/(2\sqrt{ab-c^2}),\\
\tau = -i\sqrt{b}/\sqrt{2(ab-c^2)}
\eea
\endnumparts

\noindent
Invariants:
\be
j=0,~I_1 = -2,~N= -4\sqrt{b}(\cos(s/2)-\sin(s/2))/\sqrt{2(ab-c^2)}
\ee
The Gaussian curvature of the surfaces $x_1=const$ is the negative constant $K = -b/(ab-c^2)$.

\subsection{$G_3$, Bianchi type VII (first case)} 
The commutator relations are satisfied with
\bea \nn 
\stackrel[1]{}{\xi^i}= \delta^i_3, ~\stackrel[2]{}{\xi^i}= \delta^i_2,
~\stackrel[3]{}{\xi^i}= (A[x_1], x_3+qx_2, -x_2). 
\eea   
The KE for the third KV can be written
\be  \label{q1}
AE_{,1} + 2qE -2F =0,~AF_{,1} +Fq +E -2F =0,~ AG_{,1} + 2F =0.
\ee 
In the case $A(x_1)=0$ we have from the third equation $F=0$ and from the first equation $qE=0$. Since $E \neq 0$ for a nonsingular metric, $q=0$ is required. Thus a first case of a $GVII$ metric is
\bea \label{k315}
\bold {G_3VII_{.1}:} \nn \\
E = f(x_1), ~ F = 0,~ G = f(x_1)
\eea
with a single function $f(x_1)$.
Rotation coefficients:
\be
\rho = -f_{,1}/(2f), \sigma = 0,~\nu = 0,~ \tau =0,
\ee
No invariants for $f_{,1} \neq 0$.
The surfaces $x_1 = const$ are flat, $K=0$. If $f_{,1}=0$, $K$ is invariant,
 we have the horizon \eref{hor3a}.

If $A(x_1) \neq 0$, $A=1$ can be reached with a coordinate transformation $x'_1 = f(x_1)$. The differential equations \eref{q1} are then integrated with the integration constants $a_1,a_2,a_3$:
\numparts
\bea  \label{k316}
\bold {G_3VII_{.2}:} \nn \\
E &=&  e^{-qx_1} \bigl(a_1+a_2\cos(px_1) + a_3\sin(px_1)\bigr) \\
F &=&  e^{-qx_1} \bigl(a_1 q + (a_2q+a_3p)\cos(px_1) \\
&& +(-a_2p+a_3q)\sin(px_1)\bigr)/2  \nn \\           
G &=&  e^{-qx_1}\bigl(2a_1 + (a_2[2-p^2] +a_3pq)\cos(px_1) \\
&& + (a_3[2-p^2]-a_2pq)\sin(px_1)\bigr) /2    \nn 
\eea
\endnumparts
with $p^2+q^2=4$. The determinant simplifies to
\be \label{k317}
|g_{AB}| = p^2\exp(-2qx_1) (a_1^2-a_2^2-a_3^2)/4.
\ee
We also have, using the abbreviations $\lambda = a_2(2-q^2)-a_3pq,~\mu= a_3(2-q^2)+a_2 pq$,
\be
\rho =\frac{q}{2},
~ |\sigma|=\frac{p\sqrt{a_2^2+a_3^2}}{2\sqrt{a_1^2-a_2^2-a_3^2} } ,~\tau=0,  
\ee
\numparts
\bea 
\sigma_1 &=& \frac{p(\mu\cos{px_1}-\lambda\sin{px_1})}{2(2a_1-\lambda\cos{px_1}-\mu\sin{px_1})}, \\
\sigma_2 &=& \frac{pa_1}{2\sqrt{a_1^2-a_2^2-a_3^2}}+
\frac{p\sqrt{a_1^2-a_2^2-a_3^2}}{\lambda\cos(px_1) +\mu\sin(px_1)- 2a_1}, 
\eea
\endnumparts
\be
\nu = -\frac{pa_1}{2\sqrt{a_1^2-a_2^2-a_3^2}}+~\frac{p\sqrt{a_1^2-a_2^2-a_3^2}}
{2a_1 -\lambda\cos{px_1}-\mu\sin{px_1}}.
\ee
where $\sigma =\sigma_1+i\sigma_2$.

The invariants are constant:
\numparts
\bea
j &=& \frac{q}{p}\sqrt{a_1^2/(a_2^2+a_3^2) -1}, \\
I_1 &=& - \frac{2a_1}{\sqrt{a_2^2+a_3^2}},\\
I_2 &=& 0,\\                              
I_3 &=& 16(a_1^2/(a_2^2+a_3^2)-1)/p^2, \\
M &=& 0.
\eea
\endnumparts
The surfaces $x_1 = const$ are 
flat. In the special case $a_2=a_3=0$ no invariants exist.

\subsection{$G_3$, Bianchi type VII (second case)} 
Consider now the case where the first two Killing fields are coplanar with the null direction. Here we have
\ben 
\stackrel[1]{}{\xi^i}= \delta^i_3,~ 
\stackrel[2]{}{\xi^i}= (1,0,\gamma(x_2))
\een 
satisfying $[X_1,X_3]=0$. The remaining commutator relations $[X_1,X_2]=X_2,~[X_2,X_3]= -X_1+qX_2$ involve the third KV $X_3$ and lead to
\ben     
A = x_3+(q-\gamma)x_1+a_0(x_2),~ B = b_0(x_2),~  C =\gamma x_3 + c_0(x_2)
\een
and to the condition
\be \label{con}
\gamma^2 +1 -q\gamma - \gamma_{,2}b_0(x_2) =0.
\ee
The metric is given by \eref{me1}. With the transformation \ref{tr4} we reach $a_0(x_2)=0.$ $b_0$ transforms as $b'_0(x'_2)= g_{,2}b_0(x_2)$. 
Thus either $b_0(x_2)=0$ holds or $b_0(x_2)=1$ can be reached. 
In the case $b_0=0$ one concludes from \eref{con} that (for $\gamma$ to be real) $q^2=4$ is required, which is not possible, hence $b_0=1$.  Now the relation \eref{con} can be integrated:
\be
\gamma = 
\frac{q}{2} + \frac{p}{2}\tan(\frac{p}{2}x_2).        
\ee
An integration constant was absorbed in a changed $x_2$, note also $p^2+q^2=4$. Constraining the metric
\eref{me1} with the KE for $\stackrel[3]{}{\xi^i}$ leads finally to
\numparts
\bea \label{k318}
\bold {G_3VII_{.3}:} \nn \\
E &=& a + b\frac{p^4x_1^2e^{-qx_2}}{16\cos^2(px_2/2)} -
  c\frac{p^2x_1e^{-qx_2/2}}{2\cos(px_2/2)}  \\
F &=& c e^{-qx_2/2} \cos(px_2/2)-bp^2x_1e^{-qx_2}/4 \\
G &=& b e^{-qx_2} \cos^2(px_2/2)
\eea
\endnumparts
and
\be
EG-F^2 = (ab-c^2)\cos^2(px_2/2)e^{-qx_2}.
\ee 
$a,b,c$ are constants with $ab-c^2>0$.
The rotation coefficients are
\bea
\rho=0,~\sigma= -ib\frac{p^2e^{-qx_2/2}}{8\cos(px_2/2)\sqrt{ab-c^2}} \\
\nu = b \frac{p^2e^{-qx_2/2}}{8\sqrt{ab-c^2}\cos{px_2/2}}, ~
 \tau = -i\frac{\sqrt{b}(q+p\tan(px_2/2))}{2\sqrt{2}\sqrt{ab-c^2}}.
\eea
The invariants are:
\be
j=0,~ I_1 = 2,~N=-2\sqrt{2b}\gamma(\cos(s/2)-\sin(s/2))/\sqrt{ab-c^2}.
\ee
The Gaussian curvature of the two-dimensional surfaces $x_1=const$ is
\be
K = \frac{b}{2(ab-c^2)}(2-q^2 -pq \tan(px_2/2)).
\ee

\subsection{$G_3,$ Bianchi type VIII}
%
%
The first two KV's of a $G_3VIII$ satisfy $|X_1,X_2| =X_1$, corresponding to a $G_2II$. We assume that $X_1$ and $X_2$ are non-coplanar to the null direction. The results of section \eref{g23} can be adopted: We have $\xxi{1}^i= \delta^i_3,~\xxi{2}^i= \delta^i_2+x_3\delta^i_3$ and the corresponding metric \eref{ke3}. For the third KV the communicator relations $|X_1,X_3|=2X_2,~|X_2,X_3|=X_3$ give after integration
\be \label{k319}
A=k(x_1)e^{x_2},~ B= 2x_3+ le^{x_2},~ C= x_3^2 + me^{2x_2},
\ee
where $k=k(x_1)$ depends on $x_1$, $l,m$ are constants. The KE for $X_3$ yields a system of ordinary differential equations for the functions $a, b, c$, which are present in \eref{ke3}:
\bea
a_{,1}k + 2al +4cm =0, \label{k320} \\
b_{,1}k - 2bl +4c = 0, \label{k321}\\
c_{,1}k + 2a +2bm = 0. \label{k322}
\eea
The remaining freedom in the choice of coordinates
\be
x_1' =f(x_1),~ x_2' = x_2 + g_0,~ x_3' = x_3 + h_0e^{x_2}
\ee
makes it possible to change $k, l, m$: 
\bea
k'(x_{,1}') = e^{-g_0}f_{,1}k(x_{,1}) \label{k323} \\
l' = e^{-g_0}(l - 2h_0), \label{k324}\\
m' = e^{-2g_0}(m +lh_0 -h_0^2). \label{k325}
\eea
We first assume $k(x_1) \neq 0$. By choosing $f_{,1}= e^{g_0}/k(x_1)$, we 
reach $k'(x'_1)= 1$. With $h_0= l/2$ we have $l'=0$, while $m$ changes to 
$m' = e^{-2g_0}(m + l^2/4)$. 
Depending on the sign of $m+l^2/4$, there are three cases to consider, $ m'=\pm 1,~0$.
After all these transformations have been carried out, \eref{k320}-\eref{k322} 
obtains the simplified form 
\bea
a_{,1} +4cm =0, \nn \\
b_{,1} +4c = 0,  \label{k9} \\
c_{,1} + 2a +2bm = 0, \nn 
\eea
with $m= \pm 1 , 0.$
Depending on $m$, there are three solutions to the system \eref{k9}, giving rise to three types of metrics:

\underline{$m=1$}

\bea
\bold {G_3VIII._1}: \nn \\
E = n_0+n_1e^{-4x_1}+n_2e^{4x_1}, \\
F = e^{-x_2}(n_1e^{-4x_1}-n_2e^{4x_1}),\\
G = e^{-2x_2}(-n_0 +n_1e^{-4x_1}+n_2e^{4x_1}), \\
EG-F^2 = (4n_1n_2-n_0^2)e^{-2x_2}
\eea 
$n_0,n_1,n_2, k= \sqrt{4n_1n_2-n_0^2}$ are integration constants. 
The rotation constants are
\bea
\rho=0,~ |\sigma|= 4\sqrt{n_1n_2}/k, \\
\nu =  2\frac{4n_1n_2 -n_0n_1e^{-4x_1} -n_0n_2e^{4x_1}}{k(-n_0 +n_1e^{-4x_1}+n_2e^{4x_1})}, \\
\sigma = \frac{2(n_1 e^{-4x_1}-n_2 e^{4x_1})}{-n_0+n_1e^{-4x_1}+n_2e^{4x_1}} 
        - i\nu, \\
\tau = -i \frac{\sqrt{-n_0+n_1e^{-4x_1} +n_2e^{4x_1}}}{\sqrt{2}k}.
\eea
The invariants are constant:
\bea
j=0,~ I_1= -\frac{n_0}{\sqrt{n_1n_2}}.
\eea
The surfaces $x_1=const$ have the negative Gaussian curvature
\be
K = (n_0-n_1e^{-4x_1}-n_2e^{4x_1})/k^2.
\ee

\underline{$m=0$}

\bea  \bold {G_3VIII._2}: \nn \\
E = a_0,  \nn \\ 
F =(c_0-2a_0x_1)e^{-x_2}, \\
G =(b_0 -4c_0x_1 +4a_0x_1^2)e^{-2x_2}. \nn \\ 
EG-F^2 = d_0e^{-2x_2}.
\eea
$a_0,b_0,c_0, d_0= a_0b_0-c_0^2$ are constants. The rotation coefficients are
\bea
\rho=0, |\sigma|=a_0/\sqrt{d_0} \\
\sigma = \frac{2(c_0-2a_0x_1)}{b_0-4c_0x_1+4a_0x_1^2} -i\nu, \\
\nu = \frac{a_0b_0-2c_0^2 +4a_0c_0x_1 -4a_0^2x_1^2}{\sqrt{d_0}(b_0-4c_0x_1+4a_0^2x_1^2}, \\
\tau = -i \frac{\sqrt{b_0-4c_0x_1+ 4a_0x_1^2}}{\sqrt{2}\sqrt{d_0}}.
\eea
The invariants are:          
\be
j=0,~ I_1= -2, N= -4\frac{\sqrt{4a_0x_1^2+b_0-4c_0x_1}}{\sqrt{2(a_0b_0-c_0^2)}}
 (\cos(s/2)-\sin(s/2))
\ee
The surfaces $x_1=const$ have the negative Gaussian curvature
\be K=(-b_0+4c_0x_1-4a_0x_1^2)/d_0.
\ee

\underline{$m=-1$}

\bea \bold {G_3VIII._3}: \nn \\
E = n_0 +n_1\sin(4x_1)- n_2\cos(4x_1) \nn \\
F = (n_1\cos(4x_1)+n_2\sin(4x_1) )e^{-x_2} \\
G = (n_0-n_1\sin(4x_1)+n_2\cos(4x_1))e^{2x_2} \nn \\
EG-F^2 = d_0e^{-2x_2}
\eea
$n_0,n_1,n_3,d_0= n_0^2-n_1^2-n_2^2$ are constants with $d_0 > 0$ and $n_1^2+n_2^2 \neq 0$. The rotation coefficients are
\bea
\rho=0,~ |\sigma|= 2\sqrt{n_1^2+n_2^2}/\sqrt{d_0}, \\
\sigma = \frac{2(n_1\cos(4x_1)+n_2\sin(4x_1)}{n_0-n_1\sin(4x_1)+n_2\cos(4x_1)} -i\nu, \\
\nu =-\frac{2(n_1^2+n_2^2-n_0n_1\sin(4x_1)+n_0n_2\cos(4x_1))}
    {\sqrt{d_0}(n_0-n_1\sin(4x_1)+n_2\cos(4x_1) )} \\
\tau = -i\sqrt{n_0-n_1\sin(4x_1)+n_2\cos(4x_1)}/\sqrt{2d_0}    
\eea
The invariants are constant: 
\be
j=0, ~ I_1= -2n_0/\sqrt{n_1^2+n_2^2}
\ee
The Gaussian curvature of the surfaces $x_1=const$ is negative:
\be    
K =  -(n_0-n_1\sin(4x_1) +n_2\cos(4x_1))/d_0.
\ee

In the special case $n_1=n_2=0$ divergence and shear
vanish. The metric degenerates to a horizon with a forth KV $\xxi{4}^i = \delta^i_1$:
\be  
E=n_0,~F=0.~G=n_0e^{-2x_2},
\ee 
transformable into \eref{hor3c} with $a^2= 1/n_0$.

We still have to deal with the case where $X_1$ and $X_2$ are coplanar with the null direction. Following section 5.4, we have
\be
\xxi{1}^i =\delta^i_3,~ \xxi{2}^i = \delta^i_1 + x_3\delta^i_3
\ee
for the first two KV. The metric can be written in a reduced form \eref{ke6}. 
The commutator relation $[X_1,X_3] = 2 X_2$ restricts the third KV $\xxi{3}^i = (A,B,C)$: 
\be
A=2x_3+ a(x_1,x_2),~ B= b(x_2),~C= x_3^3 + c(x_2).
\ee
The remaining relation $[X_2,X_3] = X_3$ gives $a=a_0(x_2)e^{x_1},~b=c=0$.
The KE for $X_3$ lead to $a_0=0$. Thus the metric \eref{ke6} admits a third KV $\xxi{3}^i = (2x_3,0,x_3^2)$, as already noted above.

At the beginning of this section, we assumed that the function $k(x_1)$ was not zero. The case $k(x_1)=0$ still needs to be dealt with.
From \eref{k320}-\eref{k322} we obtain $l=2c/b, m=-a/b<0$. The transformations 
\eref{k324},\eref{k325} lead with suitable $g_0,h_0$ to $l=0,~m=-1$, i.e.
to $a=b,~c=0$:
\bea \bold {G_3VIII._5}: \nn \\  
E = a(x_1), ~ F = 0,~ G = a(x_1)e^{-2x_2}  \\
EG-F^2 = a^2e^{-2x_2}. \nn
\eea
Rotation coefficients are
\be
\rho= -a_{,1}/(2a),~\sigma=0,~\nu=0,~\tau= -1/\sqrt{2a},
\ee
and the surfaces $x_1=const$ have the negative Gaussian curvature $K=-1/a.$

\subsection{$G_3$, Bianchi type IX}
The communicator relations for Bianchi type IX are:               
\ben  |X_1X_2|= X_3, ~|X_1X_3| = -X_2, ~|X_2X_3| = X_1. \een
The first two relations fix the $x_3$ dependence of the KV components:
\bea
\alpha &=& a_1 \cos(x_3) + a_2 \sin(x_3), \nn \\
\beta &=&  b_1 \cos(x_3) + b_2 \sin(x_3), \nn \\
\gamma &=& c_1 \cos(x_3) + c_2 \sin(x_3), \nn \\
A  &=& -a_1 \sin(x_3) + a_2 \cos(x_3), \nn \\
B  &=& -b_1 \sin(x_3) + b_2 \cos(x_3), \nn \\
C  &=& -c_1 \sin(x_3) + c_2 \cos(x_3). \nn 
\eea
$a_i,b_i,c_i$ generally depend on $x_1$ and $x_2$, but the first part of the KE \eref{gl2} shows that $b_i$ and $c_i$ are independent of $x_1$. The final commutator relation $|X_2X_3| = X_1$ leads to
\bea \label{a}
a_1a_{2,1} + b_1a_{2,2} -a_2a_{1,1} -b_2a_{1,2} - a_1c_1 -a_2c_2 = 0 ,\\
b_1b_{2,2} - b_2b_{1,2} -b_1c_1 - b_2c_2 = 0, \label{b} \\
b_1c_{2,2} - b_2c_{1,2} -1 -c_1^2 -c_2^2 = 0.  \label{c} 
\eea
We may simplify the last three equations with help of the coordinate transformations \eref{tr2}. Under \eref{tr2} the $a_i,b_i,c_i$ transform as
\bea
a'_1 &=& \sin(h)(f_{,1}a_2 + f_{,2}b_2) + \cos(h)(f_{,1}a_1 + f_{,2}b_1) \\
a'_2 &=& \cos(h)(f_{,1}a_2 + f_{,2}b_2) - \sin(h)(f_{,1}a_1 + f_{,2}b_1) \\
b'_1 &=& g_{,2}(b_2 \sin(h) + b_1 \cos(h)), \\
b'_2 &=& g_{,2}(b_2 \cos(h) - b_1 \sin(h)), \\
c'_1 &=& \sin(h)(c_2 + b_2h_{,2} ) + \cos(h)(c_1 + b_1h_{,2})       \\
c'_2 &=& -\sin(h)(c_1 + b_1h_{,2} ) + \cos(h)(c_2 + b_2h_{,2}).  
\eea
We first consider the transformation of $b_1$ and $b_2$. From \eref{c} follows that $b_1$ and $b_2$ cannot vanish simultaneously. Let $b_1 \neq 0$. We may reach $b'_1 =1, b'_2 =0$ by choosing $h(x_2)$ and $g(x_2)$ as
\bea \nn
\tan(h) = b_2/b_1, ~ g_{,2} = 1/\sqrt{b_1^2+b_2^2}.
\eea
The obtained values $b_1=1,~b_2=0$ are not changed for residual transformations with $h=0,~g_{,2}=1$.  One obtains from \eref{b} $c_1=0$ and from \eref{c}  
$c_{2,2} = 1 + c_2^2$. Integration gives $c_2 = \tan(x_2 + const)$ or $c_2 = \tan(x_2)$ after removing the integration constant. The still available transformations $x'_1 = f(x_1,x_2), x'_2=x_2,x'_3 = x_3$ lead to
\be
a'_1 = f_{,1} + f_{,2}, ~ a'_2 = f_{,1} a_2.
\ee
By solving the partial differential equation $f_{,1}a_1 + f_{,2} =0$ we obtain $a'_1 =0$. $a_1$ remains zero provided $f_{,2}=0$.
Note that equation \eref{a} is already satisfied if $a_2 =0$. However, a vanishing of $a_2$ cannot be reached by coordinate transformations since $f_{,1} \neq 0$. Thus the case $a_2 =0$ must be treated separately.

\underline{$a_2 = 0$}

We solve the second part of the KE \eref{gl3} for
 the metric components $E,F,G$. This gives  
\bea 
\bold {G_3IX_{.1}:}  \nn \\
E = m(x_1),~ F=0,~ G = m(x_1)\cos(x_2)^2. \label{k26}
\eea
$m(x_1)$ is an arbitrary function of $x_1$. The rotation coefficients are
\bea
\rho &=& -m_{,1}/(2 m),  \\
\sigma  &=& 0, ~ \nu = 0, \\
\tau &=& - i \tan(x_2)/\sqrt{2 m}. 
\eea

The two-dimensional wave surfaces $x_1= const$ are spheres with the Gaussian curvature $K= 1/m(x_1)$. The light cone of the FLRW cosmologies is an example for this type of null hypersurfaces. In the special case $m=const$ the metric becomes the horizon $G_\infty \times G_3IV$, \eref{hor3c}.

\underline{$a_2  \neq 0$}

If $a_{,2} \neq 0$, equation \ref{a} with $a_{1}=0$ leads to $a_{2,2} - a_2 \tan(x_2) = 0$ and after integration to $a_2 = a_{20}/\cos(x_2)$, where $a_{20}$ might depend on $x_1$. But the transformation $a'_2 = f_{,1} a_2$ finally allows to put $a_2 = 1/\cos(x_2)$.
The second part of the KE \eref{gl3} gives after integration

\bea 
\bold {G_3IX_{.2}:} \nn \\
E  =  l + m\sin(2x_1) + n\cos(2x_1), \nn \\
F  =  \cos(x_2)(n\sin(2x_1) -m\cos(2x_1)), \label{k27}\\
G  =  \cos(x_2)^2 (l - m\sin(2x_1)- n\cos(2x_1)), \nn
\eea
\be \label{k28} 
EG-F^2 =(l^2-m^2-n^2) \cos^2(x_2)   
\ee 
$l,m,n$ are integration constants here. We also note some rotation coefficients:
\bea
\rho &=& 0,  \\
|\sigma| &=& \sqrt{m^2+n^2}/\sqrt{l^2-m^2-n^2}, \\
\nu &=& \frac{n^2+m^2-l(n\cos(2x_1)+m\sin(2x_1))}{\sqrt{l^2-m^2-n^2}(l-m\sin(2x_1)-n\cos(x_1))}, \\
\tau &=& -\frac{i\tan(x_2)\sqrt{l-m\sin(x_1) -n\cos(x_1)}}{\sqrt 2 \sqrt{l^2-m^2-n^2}}
\eea
The only second-order invariant is
\be
I_1 = \frac{2l}{\sqrt{m^2+n^2}}.
\ee
The two-dimensional wave surfaces $x_1 = const$ have the Gaussian curvature:
\be
K = \frac{l -n\cos(2x_1) -m\sin(2x_1)}{l^2-m^2-n^2}.
\ee
In the case $m=n=0$ also this metric becomes the horizon metric \eref{hor3c}.

\section{Groups $G_4$ with spacelike generators}
The groups of motion of order 4 are also classified in Petrov's book \cite{petrov}. Every $G_4$ contains a subgroup $G_3$, so we can make use of previous results:
Every metric admitting a $G_3$ must be checked to see whether it permits an additional fourth KV $X_4$. $X_4$ is written as
$\xxi{4}^i = (U,~V,~W)$. The first part of the KE requires  
$V_{,1}= 0,~W_{,1}=0$.

\subsection{Group  $G_{4I}$}
The group operators satisfy the relations
\bea
|X_1,X_2| =0,~ |X_2,X_3| = X_1,~ |X_3,X_1| =0. \\
|X_1,X_4| = kX_1, ~|X_2,X_4| = X_2,~|X_3,X_4| = (k-1)X_3.
\eea
The subgroup is $G_{3II}$, hence metric \eref{k31} applies. 
The commutation relations with $X_4$ lead to 
\be
U= (k-1)x_1 = const,~ V= x_2,~ W = kx_3 = const.
\ee
One KE for $X_4$ gives $bk=0$, thus $k=0$. Another KE then leads to $ab-c^2=0$, corresponding to a singular metric.
No compatible metric exists.

\subsection{Group  $G_{4II}$}
Commutator relations:   
\bea
|X_1,X_2| =0,~ |X_2,X_3| = X_1,~ |X_3,X_1| =0, \\
|X_1,X_4| = 2X_1, ~|X_2,X_4| = X_2,~|X_3,X_4| = X_2+X_3.
\eea
Again $G_{3II}$ is subgroup, also here the KE for $X_4$ is not compatible 
with the commutator relation $|X_3,X_4| = X_2+X_3$.
No compatible metric exists.
 
\subsection{Group  $G_{4III}$}
Commutator relations: 
\bea
|X_1,X_2| =0,~ |X_2,X_3| = X_1,~ |X_3,X_1| =0, \\
|X_1,X_4| = qX_1, ~|X_2,X_4| = X_3,~|X_3,X_4| = -X_2+qX_3.
\eea
Also here a $G_{3II}$ is subgroup, and the commutator relations 
for $X_4$ cannot be satisfied.
No compatible metric exists.

\subsection{Group  $G_{4IV}$}
The commutator relations:
\bea
|X_1,X_2| =0,~ |X_1,X_3| = X_1,~ |X_2,X_3| =0, \\
|X_1,X_4| = 0, ~|X_2,X_4| = X_2,~|X_3,X_4| = 0  
\eea
\emph{differ from those given by Petrov}, since we have interchanged 
$X_4$ und $X_3$. Thus $G_{3III}$ becomes a subgroup of $G_{4IV}$. 
Two metrics, \eref{k33} and \eref{k39} belong to  $G_{3III}$. The first becomes singular, if the KE for $X_4$ is taken into account. For the second metric \eref{k39} the commutator relations for $X_4$ give
\be U= x_1+ f(x_3),~V = const,~ W= const~ e^{x_2},~ k(V+1) = 0.
\ee
The KE for $X_4$ leads to $bV=0$ and $bW_2 =0$, hence (since $b \neq 0$) to $V=0,~W=0$ and $k=0$. 
The metric is reduced to
\be
E = a,~ F = c e^{-x_2},~ G = b e^{-2x_2} 
\ee
and represents a horizon, equivalent to \eref{hor3b}. $X_2$ becomes a null vector. The fourth KV $X_4$ is also null and can be written  $\xxi{4}^i = (x_1,0,0)$.

\subsection{Group  $G_{4V}$}
Commutator relations:   
\bea
|X_1,X_2| =0,~ |X_2,X_3| = X_1,~ |X_3,X_1| =0, \\
|X_1,X_4| = 2X_1, ~|X_2,X_4| = X_2,~|X_3,X_4| = X_2+X_3.
\eea
Subgroup is $G_{3V}$. Two metrics are admitted, \eref{k35} and \eref{k311}.
For the first metric the commutator relations for $X_4$ give
\be
U = const,~ V= x_3,~ W = -x_2
\ee
and the KE reads
\be
c-aU= 0,~ b-a +2cU = 0,~ c- bU =0.
\ee
This leads to $c=0, ~a=b,~ U=0$. the metric becomes
\bea
\bold {G_{4V.1}:} \nn \\
E = a e^{-2x_1},~ F = 0,~ G = a e^{-2x_1}.
\eea
\be
EG-F^2 = a^2 e^{-4x_1}.
\ee

The rotation coefficients are
\be 
\rho = 1,~\sigma =0,~\nu = 0,~\tau = 0
\ee
\noindent
There are no invariants, the surfaces $x_1=const$ are flat.

The other metric \eref{k311} is not compatible with the KE for $X_4$.  

\subsection{Group  $G_{4VI}$}
The commutator relations are
\bea
|X_i,X_j| =0,~ |X_i,X_4| = C_i^k X_k,\\ 
i,j =1,2,3;~ k=1,2,3,4
\eea
$G_4VI$ contains a $G_3I$ group as subgroup. Since all $G_3I$ metrics are 
horizons, this also holds for $G_4VI$ metrics.

\subsection{Group  $G_{4VII}$}
The commutator relations are
\bea
|X_1,X_2| = X_1, ~|X_1,X_3| = 2 X_2,~ |X_2,X_3| =  X_3, \\
|X_i,X_4| = 0. ~ i=1,2,3. 
\eea
The group $G_{4VII}$ contains a subgroup $G_{3VIII}$. Five metrics satisfy 
the symmetries of $G_{3VIII}$, but only $G_{3VIII._4}$ admits a space-like KV 
$X_4= \partial_2$ that satisfies the commutator relations. The KE for $X_4$ 
require $k=const>1$:
\bea 
\bold {G_{4VII}:} \nn  \\
E = 1,~ F=e^{-x_1},~ G = ke^{-2x_1}.   
\eea
The rotation coefficients are
\be
\rho=\frac{1}{2},~ \sigma=\frac{1}{2}+\frac{i}{2\sqrt{k-1}},~\nu=-\frac{1}{2\sqrt{k-1}},~ \tau =0.
\ee
Invariants are 
\be
j= \sqrt{1-1/k},~I_1= -\frac{2}{\sqrt{k}}, I_2=0,~M=0.
\ee
The Gaussian curvature of the $x_1=const$ surfaces vanishes.

\subsection{Group  $G_{4VIII}$}

The commutator relations are
\bea
|X_1,X_2| = X_3, ~ |X_1,X_3| = -X_2,~ |X_2,X_3| = X_1. \\
|X_i,X_4| = 0. ~ i=1,2,3. 
\eea
$G_{4VIII}$ has $G_{3IX}$ as a subgroup and the associated metrics 
are \eref{k26} and \eref{k28}. For \eref{k26}, the commutator relations 
yield $U=f(x_1), V=0, W=0$. The KE limits the metric by $m_{,1} = 0$, 
thus, with constant $m$,
\be 
E = m,~ F=0,~ G = m\cos(x_2)^2.
\ee
This is the metric of the horizon \eref{hor2}.
If we start from the second metric \eref{k28}, commutator relations and KE lead 
to the same final metric.

\section{Final comments} 
A group $G_r$ is called transitive, if any  two points on the manifold can be reached by group orbits. A well-known theorem \cite{eisenhart2} states: A necessary and sufficient condition that a group $G_r$ be transitive is that $r \geq n$ and that the rank of the matrix $(\xxi{s}^i)~ (i=1 ...n,s=1 ...r)$ is $n$. Since $n=3$ in our case, this applies only to the groups $G_3$ and $G_4$, the groups $G_1$ and $G_2$ are intransitive. The rank condition identifies the following $G_3$ metrics as intransitive:
\bea
\bold{G_3VII._1}, q=0, \\
\bold{G_3VIII._5}
\eea
Both manifolds are covered by two-dimensional surfaces $x_1=const$, on which the metric is transitive.

The set of all points a given point $x_i$ is mapped by the group operators of a $r$-parameter group $G_r$ is called the \emph{orbit} of $x_i$.  
The orbits of the metrics discussed here can be determined relatively easily by integration. But this topic belongs to global issues that are not covered here.          

\newpage

\section*{Appendix A:~ Classification of null hypersurfaces in terms of invariants}

Geometrical properties of null hypersurfaces can be expressed in terms of differential invariants, as described in \cite{daut2,daut3}. The differential invariants -
 up to second order - are found in section 3 as triad-invariant functions of the rotation coefficients $\rho$ (divergence), $\sigma$ (shear) and $\tau,~  \nu,~ \chi$ and $\varphi$.  
Equivalently, the invariants are also uniquely determined by the inner metric $\gamma_{ik}$. Written as functions of the rotation coefficients, there is considerable freedom in their representation, depending on the choice of the coefficients.
In this paper we have used adapted coordinates with $\gamma_{1A}=0$. A possible representation of rotation coefficients as functions of the inner metric in adapted coordinates is given below, together with the differential invariants up to second order. A classification of null hypersurfaces along these lines is given in Table 1.  

\begin{table}
\small        
\caption{\label{AppB}Classification of null hypersurfaces in term of invariants.}           
\lineup
	\begin{tabular}[h]{@{}*{5}{|l|l}}  
\br
shear & divergence & conditions                    & class &  invariants \cr
\mr	
$|\sigma| \neq 0$ & $\rho \neq 0$ &$I_2 \neq 0$ & $1$ & $j,I_1,I_2,J$ \cr
&   & $I_2 =0,~I_1^2-4(1-j^2) \neq 0$ & $2a$ & $j,I_1,I_2=0,J,M  $    \cr
&   & $I_2 =0,~I_1^2-4(1-j^2) =0    $ & $2b$ & $j,I_1,I_2=0,J=0,L,M$    \cr
		\cline{2-5}		
&  $\rho =0$ & $|I_1| = 2$    & $3$ & $j=0,~|I_1|=2,~N$   \cr
&            & $|I_1| \neq 2$ & $4$ & $j=0,~I_1$ \cr
\mr 
$|\sigma| =0$       & $\rho \neq 0$ &          & $5$& $no~invariants$ \cr
	& $\rho = 0$ &  & $6$ & $K$ \cr
	\mr	
\end{tabular}
\end{table}

With $\Gamma = |\gamma_{AB}|$ and the usual abbreviations $E=\gamma_{22},~F=\gamma_{23},~ G=\gamma_{33}$  we have:
\bea
\fl \rho = -\Gamma_{,1}/(4\Gamma), \\
\fl \sigma = \Gamma_{,1}/(4\Gamma) - G_{,1}/(2G) + \frac{i}{2\sqrt{\Gamma}}(F_{,1}-FG_{,1}/G), \\
\fl \nu =(G_{,1}F-G F_{,1})/(2G\sqrt{\Gamma}),\\
\fl \tau = \frac{i}{2\sqrt{2G}}(G_{,3}/G - \Gamma_{,3}/\Gamma) + 
\frac{i}{2\sqrt{2\Gamma G}}(G_{,2}+ FG_{,3}/G- 2F_{,3}), \\
\fl \chi = \varphi = 0.
\eea
The differential operators are given by
\be
D = \partial_1, ~\delta= \frac{i\sqrt{G}}{\sqrt{2\Gamma}}\partial_2 + \frac{\sqrt{\Gamma} - iF}{\sqrt{2\Gamma G}} \partial_3.
\ee
The differential invariants in Table 1 are defined in terms of rotation coefficients by (notice $e^{is} = \sigma/|\sigma|$ and the abbreviations $\zeta=2j-iI_1+I_2,~\vartheta= 2j-iI_1-I_2$)
\bea
j = \rho/|\sigma|, \\
I = I_1+iI_2 = \frac{i}{|\sigma|}\Bigl(\frac{D\rho}{\rho}- 
\frac{D\sigma}{\sigma}\Bigr) +\frac{2\nu}{|\sigma|}, \\
J = e^{is/2}\Bigl[ (\vartheta\bar{\vartheta}-4)\Bigl(\frac{\delta \sigma}{\sigma}
-\frac{\delta  \rho}{\rho} \Bigr) +(\zeta\bar{\vartheta}
-4)\Bigl(\frac{\delta \bar{\sigma}}{\bar{\sigma}}
-\frac{\delta \rho}{\rho}\Bigr) +2\bar{\vartheta}(\zeta-\vartheta)\bar{\tau}
\Bigr] \nn \\
\qquad +e^{-is/2}\Bigl[2(\zeta-\vartheta)\Bigl(\frac{\bar{\delta}\sigma}{\sigma} -\frac{\bar{\delta}\rho}{\rho} +2\tau \Bigr) \Bigr],  \\
K = \delta \tau +\bar{\delta} \bar{\tau} -2\tau\bar{\tau}, \\
L = e^{is/2}\Bigl[\frac{\delta\sigma}{\sigma}-\frac{\delta\rho}{\rho}
-2\bar{\tau}\Bigr]+ 
e^{-is/2}(j+i\sqrt{1-j^2})\Bigl[\frac{\bar{\delta}\bar{\sigma}}{\bar{\sigma}}
-\frac{\bar{\delta}\rho}{\rho} -2\tau\Bigr],
\\
M = -2e^{is/2} \delta j/j,  \\
N= e^{is/2}(\frac{\delta \sigma}{\sigma}-\frac{\delta \bar{\sigma}}{\bar{\sigma}}-4\bar{\tau})/(1+i)
-ie^{-is/2}( \frac{\bar{\delta} \sigma}{\sigma}
 -\frac{\bar{\delta} \bar{\sigma} }{\bar{\sigma}} +4\tau)/(1+i). \eea
We stress again: $j,I,J,K,L,M,N$ are invariants only, if certain conditions are satisfied for the rotation coefficients. These conditions can be taken from Table 1. E.g., $N$ is an invariant if and only if $|\sigma| \neq 0,~ \rho = 0$ and $|I_{1}|=2$.

\section*{Appendix B:~ Overview of metrics with Killing 
symmetries}  

\begin{table}[h]
\small        
\caption{\label{1} Group $G_\infty$ (horizon) }
\lineup
\begin{tabular}{@{}*{4}{|l|l}}  
\br
metric\ & generators\ &  invariants\ & class \cr
\mr	
$E= E(x_2,x_3)$ & $X_0=f(x_1)\partial_1$ & Gaussian     & 6 \cr
$F= F(x_2,x_3)$ &                   & curvature $K$& \cr
$G= G(x_2,x_3)$ &                   &              &\cr
\mr     
\end{tabular}
\end{table}

\begin{table}[h]
\small        
\caption{\label{2} Group $G_\infty \times G_1I$ (horizon) }
\lineup
\begin{tabular}{@{}*{4}{|l|l}}  
\br
metric\ & generators\ &  invariants\ & class \cr
\mr	
$E= 1         $ & $X_0 =f(x_1)\partial_1$ & $K=$        & 6 \cr
$F= 0         $ & $X_1 =\partial_3$ & $(-2mm_{,22}+m_{,2}^2)/m^2$  & \cr
$G= m(x_2)    $     &                   &              &\cr
\mr     
\end{tabular}
\end{table}

\begin{table}[h]
\small        
\caption{\label{3a} Group $G_\infty \times G_3VII_0$~(horizon) }
\lineup
\begin{tabular}{@{}*{4}{|l|l}}  
\br
metric\ & generators\ &  invariants\ & class \cr
\mr	
$E= a^2    $ & $X_0 =f(x_1)\partial_1$ & $K=0$        & 6 \cr
$F= 0      $ & $X_1 =\partial_3, X_2=\partial_2$ &   & \cr
$G= a^2    $ & $X_3 =x_3\partial_2-x_2\partial_3$ &   &\cr
$a=const$ & & & \cr
\mr     
\end{tabular}
\end{table}

\begin{table}[h]
\small        
\caption{\label{3b} Group $G_\infty \times G_3VIII$ (horizon) }
\lineup
\begin{tabular}{@{}*{4}{|l|l}}  
\br
metric\ & generators\ &  invariants\ & class \cr
\mr	
$E= a^2/x_2^2   $ & $X_0 =f(x_1)\partial_1$ & $K=-1/a^2$        & 6 \cr
$F= 0   $ & $X_1 =\partial_3,~X_2=x_2\partial_2+x_3\partial_3$ &  & \cr
$G= a^2/x_2^2 $  & $X_3 = 2x_2x_3\partial_2+ (x_3^2-x_2^2)\partial_3 $  &  &\cr
$a=const$ & & & \cr
\mr     
\end{tabular}
\end{table}

\begin{table}[h]
\small        
\caption{\label{3c} Group $G_\infty \times G_3IX$ (horizon) }
\lineup
\begin{tabular}{@{}*{4}{|l|l}}  
\br
metric\ & generators\ &  invariants\ & class \cr
\mr	
$E= a^2  $ & $X_0 =f(x_1)\partial_1$ & $K=1/a^2$        & 6 \cr
$F= 0 $ & $X_1 =\partial_3,~X_2=\cos{x_3}\partial_2+
sin{x_3}\tan{x_2}\partial_3  $   &  & \cr
$G= a^2\cos{x_2}^2  $  & 
$X_3=-\sin{x_3}\partial_2+\cos{x_3}\tan{x_2}\partial_3$ &   &\cr
$a=const$ & & & \cr
\mr     
\end{tabular}
\end{table}

\begin{table}[h]
\small        
\caption{\label{4} Group $G_1$ }
\lineup
\begin{tabular}{@{}*{4}{|l|l}}  
\br
metric\ & generators\ &  invariants\ & class \cr
\mr	
$E= E(x_1,x_2)$ & $X_1 =\partial_3$  & $j,I,J$      &  1        \cr
$F= F(x_1,x_2)$ &                    & $(see App. A)$ & $(in$      \cr
$G= G(x_1,x_2)$ &                    &              & $general)$  \cr
\mr     
\end{tabular}
\end{table}

\begin{table}[h]
\small        
\caption{\label{5} Abelian Group $G_2I.1$~ \\(space-like generators $X_1,X_2$, not coplanar with null direction)}
\lineup
\begin{tabular}{@{}*{4}{|l|l}}  
\br
metric\ & generators\ &  invariants\ & class \cr
\mr	
$E= a(x_1)$ & $X_1 =\partial_3$   &$j=-d_{,1}/\sqrt{p}$   & 1 \cr
$F= c(x_1)$  & $X_2 = \partial_2$ & $ I_1 =  8(d/p)^{3/2}(a_{,11}[b_{,1}c-bc_{,1}]   $  &  \cr
$G= b(x_1)$ &   &$~~~~~~ +b_{,11}[c_{,1}a-a_{,1}c]+c_{,11}[a_{,1}b-b_{,1}a]) $&\cr
	$d=ab-c^2$ &       &$I_2= -4dj_{,1}/d_{,1}    $   &  \cr
$p= (a_{,1}b-ab_{,1})^2+ $&   &$J = 0$   & \cr
$~4(a_{,1}c-ac_{,1})(b_{,1}c-bc_{,1})\neq0$ &  &  & \cr
\mr     
\end{tabular}
\end{table}

\begin{table}[h]
\small        
\caption{\label{$G_2.2$} Abelian Group $G_2I.2$~ \\(space-like generators $X_1,X_2$, not coplanar with null direction)}
\lineup
\begin{tabular}{@{}*{4}{|l|l}}  
\br
metric\ & generators\ &  invariants\ & class \cr
\mr	
$E= a(x_1)$ & $X_1 =\partial_3$   &    & 5 \cr
$F= c(x_1)$  & $X_2 = \partial_2$ &    &  \cr 
$G= b(x_1)$ &   &         &               \cr   
$p= (a_{,1}b-ab_{,1})^2 + $&   &          & \cr 
$~4(a_{,1}c-ac_{,1})(b_{,1}c-bc_{,1})=0$ &  &  & \cr
\mr     
\end{tabular}
\end{table}

\begin{table}[h]
\small        
\caption{\label{$G_2.3$} Abelian Group $G_2I.3$~ \\(space-like generators $X_1,X_2$, coplanar with null direction)}
\lineup
\begin{tabular}{@{}*{4}{|l|l}}  
\br
metric\ & generators\ &  invariants\ & class \cr
\mr	
$E= 1+x_1^2\gamma_{,2}^2m(x_2)$ & $X_1 =\partial_3$   &$j=0$   & 3 \cr
$F= x_1\gamma_{,2}m(x_2)$  & $X_2 = \partial_1 +\gamma\partial_3$ & $ I_1 = -2\gamma_{,2}/|\gamma_{,2}|   $  &  \cr
$G= m(x_2)$ & & $N = (\cos(s)-sin(s))m,_2/(\sqrt{2}m)$    &\cr
$\gamma_{,2} \neq 0 $ &       &    &  \cr
\mr     
\end{tabular}
\end{table}

\begin{table}[h]
\small        
\caption{\label{$G_2II.1$} Non-Abelian Group $G_2II.1$~ \\(space-like generators $X_1,X_2$, not coplanar with null direction)}
\lineup
\begin{tabular}{@{}*{4}{|l|l}}  
\br
metric\ & generators\ &  invariants\ & class \cr
\mr	
$E= a(x_1)$ & $X_1 =\partial_3$   & $j=-d_{,1}/\sqrt{p}$  & 1 \cr
$F= c(x_1)e^{-x_2}$  & $X_2 = \partial_2 +x_3\partial_3$ &$I_1 =  8(d/p)^{3/2}(a_{,11}[b_{,1}c-bc_{,1}] $  & \cr
$G= b(x_1)e^{-2x_2}$ & &$+b_{,11}[c_{,1}a-ca_{,1}]+c_{,11}[a_{,1}b-b_{,1}a])$& \cr     
$d=ab-c^2  $ &  & $I_2= 2\frac{d}{\sqrt{p}}(2\frac{d_{,11}}{d_{,1}}- \frac{p_{,1}}{p}) $   &  \cr
$p= (a_{,1}b-ab_{,1})^2 $&  &    &  \cr 
$~+ 4(a_{,1}c-ac_{,1})(b_{,1}c-bc_{,1})$ & &  $J= ~ see~ \eref{ke4a} $   & \cr
$d_{,1} \neq 0$  &  &  & \cr
\mr
\end{tabular} 
\end{table}

\begin{table}[h]
\small        
\caption{\label{$G_2II.2$} Non-Abelian Group $G_2II.2$~ \\(space-like generators $X_1,X_2$, not coplanar with null direction)}
\lineup
\begin{tabular}{@{}*{4}{|l|l}}  
\br
metric\ & generators\ &  invariants\ & class \cr
\mr	
$E= a(x_1)$ & $X_1 =\partial_3$   & $j = 0$ & 4 \cr
$F= c(x_1)e^{-x_2}$  & $X_2 = \partial_2 +x_3\partial_3$ & $ I_1 = -sgn(c) \frac{2B}{\sqrt{d} A^{3/2}} $ & \cr
$G= b(x_1)e^{-2x_2}$ & &$ A=(a_{,1}b+b_{,1}a)^2-4a_{,1}b_{,1}c^2 $  & \cr       
$d=ab-c^2,~ d_{,1}=0  $ &  & $B= 4dc^2(a_{,11}b_{,1}-b_{,11}a_{,1}) $ &  \cr
&  &	$~~~~+ba_{,1}^2(a_{,1}b^2+b_{,1}ab-4b_{,1}c^2) $  & \cr 
&  &    $~~~~-ab_{,1}^2(b_{,1}a^2+a_{,1}ab-4a_{,1}c^2) $  & \cr 
\mr     
\end{tabular}
\end{table} 

\begin{table}[h]
\small        
\caption{\label{$G_2II.3$} Non-Abelian Group $G_2II.3$~ \\(space-like generators $X_1,X_2$, not coplanar with null direction)}
\lineup
\begin{tabular}{@{}*{4}{|l|l}}  
\br
metric\ & generators\ &  invariants\ & class \cr
\mr	
$E= k_1b(x_1)$ & $X_1 =\partial_3$   &                & 5 \cr
$F= k_2b(x_1)e^{-x_2}$  & $X_2 = \partial_2 +x_3\partial_3$ &  & \cr
$G= b(x_1)e^{-2x_2}$ &  &  & \cr 
$b_{,1} \neq 0, k_1>k_2^2  $ &  & &  \cr
\mr     
\end{tabular}
\end{table}

\begin{table}[h]
\small        
\caption{\label{$G_3II.1$} Bianchi type $G_3$II.1}
\lineup
\begin{tabular}{@{}*{4}{|l|l}}  
\br
metric & generators & invariants & class\cr
\mr	
$E= a -2cx_1+bx_1^2$ & $X_1 =\partial_3$                & $j=0   $     & 3\cr
$F= c -bx_1$  & $X_2 = \partial_2$                      & $I_1 =2$    & \cr
$G= b$ &  $X_3 = \partial_1 +x_2\partial_3$  &$N=0$ & \cr
$a,b,c = const$&       &  &  \cr
\mr     
\end{tabular}
\end{table}

\begin{table}[h]
\small        
\caption{\label{$G_3III.1$} Bianchi type $G_3$III.1}
\lineup
\begin{tabular}{@{}*{4}{|l|l}}  
\br
metric & generators & invariants & class\cr
\mr	
$E= a $ & $X_1 =\partial_3$ & $j=\sqrt{1-c^2/(ab)}  $     & 2b\cr
$F= c e^{-x_1}$  & $X_2 = \partial_2$ & $I_1 = -2c/\sqrt{ab},~I_2=0,~I_3=0$ & \cr
$G= be^{-2x_1}$ &  $X_3 = \partial_1 +x_3\partial_3$ &$J=0,~L=0,~M=0$ \cr
$a,b,c = const$&       &  &  \cr 
\mr    
\end{tabular}
\end{table}

\begin{table}[h]
\small        
\caption{\label{$G_3III.2$} Bianchi type $G_3$III.2}
\lineup  
\begin{tabular}{@{}*{4}{|l|l}}  
\br
metric & generators & invariants & class\cr
\mr	
$E= a-2ckx_1+ bk^2x_1^2 $ & $X_1 =\partial_3$ & $j=0 $     & 3\cr
$F=(c-bkx_1)e^{-x_2} $  & $X_2 = \partial_1+ke^{x_2}\partial_3$ & $I_1 = 2sgn(k) $ & \cr
$G= be^{-2x_2}$ &  $X_3 = \partial_2 +x_3\partial_3$ & $N= 4\sqrt{b}(\cos(s/2)-\sin(s/2))/  $& \cr
$a,b,c,k = const$&       &$~~~~\sqrt{2(ab-c^2)} $  &  \cr
\mr     
\end{tabular}
\end{table}


\begin{table}[h]
\small        
\caption{\label{$G_3IV.1$} Bianchi type $G_3$IV.1}
\lineup
\begin{tabular}{@{}*{4}{|l|l}}  
\br
metric & generators & invariants & class\cr
\mr	
$E=(a -2cx_1+bx_1^2)e^{-2x_1}$ & $X_1 =\partial_3$ & $j=2\sqrt{ab-c^2}/b$ & 
2a \cr 
%
$F=(c -bx_1)e^{-2x_1}$  & $X_2 = \partial_2$ & $I_1 =2,~I_2=0~$    & \cr
$G=be^{-2x_1}$ &  $X_3 = \partial_1 +x_2\partial_2 + (x_2+x_3)\partial_3$ &$I_3=16(ab-c^2)/b^2$ & \cr
$a,b,c = const$ &       &$J=0,~M = 0$  &  \cr
\mr     
\end{tabular}
\end{table}

\begin{table}[h]
\small        
\caption{\label{$G_3IV.2$} Bianchi type $G_3$IV.2}
\lineup
\begin{tabular}{@{}*{4}{|l|l}}  
\br
metric & generators & invariants & class\cr
\mr	
$E=a+ 2cx_1e^{-x_2}+bx_1^2e^{-2x_2}$ & $X_1 =\partial_3$ & $j=0$& 3 \cr
$F=ce^{-x_2} +bx_1e^{-2x_2}$  & $X_2 = \partial_1-x_2\partial_3$ & $I_1 =-2$    & \cr
$G=be^{-2x_2}$ &  $X_3 = x_1\partial_1 +\partial_2        $ &$
	N = -4\sqrt{b}(\cos(s/2)-\sin(s/2))                 $	& \cr
$a,b,c = const$ & $~~~~+x_3\partial_3$      &$~~/\sqrt{2(ab-c^2)}$  & 
\cr
\mr     
\end{tabular}
\end{table}

\begin{table}[h]
\small        
\caption{\label{$G_3V.1$} Bianchi type $G_3$V.1}
\lineup
\begin{tabular}{@{}*{4}{|l|l}}
\br
metric & generators & invariants & class\cr
\mr	
$E= ae^{-2x_1}$& $X_1 =\partial_3$     &      & 5\cr
$F= ce^{-2x_1}$& $X_2 =\partial_2$   &    & \cr
$G= be^{-2x_1}$& $X_3 =\partial_1 +x_2\partial_2+x_3\partial_3$  & & \cr
$a,b,c = const$ &      &  &  \cr
\mr     
\end{tabular}
\end{table}

\begin{table}[h]
\small        
\caption{\label{$G_3VI.1$} Bianchi type $G_3$VI.1}
\lineup
\begin{tabular}{@{}*{4}{|l|l}}  
\br
metric & generators & invariants & class\cr
\mr
$E=ae^{-2qx_1}$ & $X_1 =\partial_3$ & $j=\frac{q+1}{q-1}\sqrt{1-c^2/(ab)} 
$ & 1\cr
$F=ce^{-(1+q)x_1} $  & $X_2 = \partial_2$ & $I_1 =2/\sqrt{ab},~ I_2=0$    & \cr
$G=be^{-2x_1}$ &  $X_3 = \partial_1 +qx_2\partial_2+x_3\partial_3$  &$I_3=16q(ab-c^2)/(ab(q-1)^2),~ J=0$ & \cr
$q \neq 0,1$&       &  &  \cr
$a,b,c = const$ &      &  &  \cr
\mr     
\end{tabular}
\end{table}

\begin{table}[h]
\small        
\caption{\label{$G_3VI.2$} Bianchi type $G_3$VI.2}
\lineup
\begin{tabular}{@{}*{4}{|l|l}}  
\br
metric & generators & invariants & class\cr
\mr	
$E=a-2ck(1-q)x_1e^{-qx_2}+ $ 
& $X_1 =\partial_3$ & $j=0$  & 3\cr
$~~bk^2(1-q)^2x_1^2e^{-2qx_2}$ &  &  &  \cr
$F=ce^{-x_2}-bk(1-q)x_1e^{-(1+q)x_2} $ & $X_2 = \partial_1 +ke^{(1-q)x_2}\partial_3$ & $I_1 = 2\frac{k(1-q)}{|k(1-q)|} $    & \cr
	$G=be^{-2x_2}$ &  $X_3 =qx_1 \partial_1 +\partial_2+x_3\partial_3$  &$N= -4\sqrt{b}(\cos(s/2) $ & \cr
	$q \neq 0,1,~ a,b,c = const$ &  & $ -\sin(s/2))\sqrt{2(ab-c^2)}$ 
& \cr
\mr     
\end{tabular}
\end{table}

\begin{table}[h]
\small        
\caption{\label{$G_3VII.1$} Bianchi type $G_3$VII.1(intransitive group)}
\lineup
\begin{tabular}{@{}*{4}{|l|l}}  
\br
metric & generators & invariants & class\cr
\mr	
$E= f(x_1)$ & $X_1 =\partial_3$ &    & 5\cr
$F= 0$  & $X_2 = \partial_2$  &     & \cr
$G= f(x_1)$ &  $X_3 = x_3\partial_2 -x_2\partial_3$  & & \cr
$q=0, f_{,1} \neq 0$	&   &   & \cr
\mr     
\end{tabular}
\end{table}

\begin{table}[h]
\small        
\caption{\label{$G_3VII.2$} Bianchi type $G_3$VII.2}
\lineup
\begin{tabular}{@{}*{4}{|l|l}}  
\br
metric & generators & invariants & class\cr
\mr	
$E = e^{-qx_1} \bigl(a_1+a_2\cos(px_1) + a_3\sin(px_1)\bigr)$ &
	$X_1 =\partial_3$    & $j=  $     & 2a\cr
$F = e^{-qx_1} \bigl(a_1 q + (a_2q+a_3p)\cos(px_1) $ &
    $X_2 = \partial_2$ &$\frac{q}{p}\sqrt{a_1^2/(a_2^2+a_3^2)-1} $  & \cr
$~~~~ +(-a_2p+a_3q)\sin(px_1)\bigr)/2 $ &   &  $I_1 =-2a_1/\sqrt{a_2^2+a_3^2}$    &    \cr
$G = e^{-qx_1}\bigl(2a_1 + (a_2[2-p^2] +a_3pq)\cos(px_1) $ & 
    $X_3 = \partial_1$ &$I_2=0   $ & \cr
$~~~~ + (a_3[2-p^2]-a_2pq)\sin(px_1)\bigr) /2$  & 
$~~+(x_3+qx_2)\partial_2  $ & $I_3=16(a_1^2/(a_2^2+a_3^2)$ &  \cr
$p,q,a_1,a_2,a_3 = const,~p^2+q^2=4$ &$~~-x_2\partial_3$ &$-1)/p^2,~J=0,~M=0$ &  \cr
\mr     
\end{tabular}
\end{table}

\begin{table}[h]
\small        
\caption{\label{$G_3VII.3$} Bianchi type $G_3$VII.3}
\lineup
\begin{tabular}{@{}*{4}{|l|l}}  
\br
metric & generators & invariants & class\cr
\mr	
$E = a + b\frac{p^4x_1^2e^{-qx_2}}{16\cos^2(px_2/2)} -
  c\frac{p^2x_1e^{-qx_2/2}}{2\cos(px_2/2)}$ & $X_1 =\partial_3$                & $j=0   $     & 3\cr
	$F= c e^{-qx_2/2} \cos(px_2/2)-bp^2x_1e^{-qx_2}/4  $  & $X_2 = \partial_1 + \gamma\partial_3$       & $I_1 =2,~N = -2\sqrt{2b}\gamma$    & \cr
$G = b e^{-qx_2} \cos^2(px_2/2)$ &  $X_3 = (x_3+(q-\gamma))\partial_1 $  &
$~~(\cos(s/2)-\sin(s/2))  $ & \cr
	$\gamma = \frac{q}{2} + \frac{p}{2}\tan(\frac{p}{2}x_2),~ p^2+q^2=4$  & $~~~+\partial_2 +x_3\gamma\partial_3$    &$~~/\sqrt{ab-c^2}$     &   \cr
\mr     
\end{tabular}
\end{table}

\begin{table}[h]
\small   
\caption{\label{$G_3VIII.1$} Bianchi type $G_3$VIII.1}
\lineup
\begin{tabular}{@{}*{4}{|l|l}}  
\br
metric & generators & invariants & class\cr
&       &  &  \cr
\mr	
$E= n_0+n_1e^{-4x_1}+n_2e^{4x_1} $ & $X_1 =\partial_3$ & $j=0$   & 4\cr
$F=e^{-x_2}(n_1e^{-4x_1}-n_2e^{4x_1})$ & $X_2 =\partial_2+x_3\partial_3$ &
	$I_1=-\frac{n_0}{\sqrt{n_1n_2}} $& \cr
$G=e^{-2x_2}(-n_0+n_1e^{-4x_1}+n_2e^{4x_1}) $ &  $X_3 =e^{x_2}\partial_1+ 
	 2x_3\partial_2 +(x_3^2+e^{2x_2})\partial_3 $  & & \cr
$n_0,n_1,n_2 = const$& & &  \cr
$ 4n_1n_2-n_0^2 >0$ &  &  & \cr
\mr     
\end{tabular}
\end{table}

\begin{table}[h]
\small   
\caption{\label{$G_3VIII.2$} Bianchi type $G_3$VIII.2}
\lineup
\begin{tabular}{@{}*{4}{|l|l}}  
\br
metric & generators & invariants & class\cr
&       &  &  \cr
\mr	
$E= a_0 $ & $X_1 =\partial_3$ & $j=0$   & 3\cr
$F=(c_0-2a_0x_1)e^{-x_2} $ & $X_2 =\partial_2+x_3\partial_3$ &
	$I_1=-2,~ N = $& \cr
$G=(b_0-4c_0x_1+4a_0x_1^2)e^{-2x_2}$ &  $X_3 =e^{x_2}\partial_1+ 
$  &$-\frac{4}{\sqrt{2}\sqrt{a_0b_0-c_0^2}}(\cos(s/2)-\sin(s/2))$ & \cr
$a_0,b_0,c_0 = const$&$~~~ 2x_3\partial_2 +x_3^2\partial_3 $ &$~~~\sqrt{4a_0x_1^2+b_0-4c_0x_1} $ &  \cr
$a_0b_0-c_0^2 >0$ &  &  & \cr
\mr     
\end{tabular}
\end{table}

\begin{table}[h]
\small   
\caption{\label{$G_3VIII.3$} Bianchi type $G_3$VIII.3}
\lineup
\begin{tabular}{@{}*{4}{|l|l}}  
\br
metric & generators & invariants & class\cr
&       &  &  \cr
\mr	
$E= n_0+n_1\sin(4x_1)-n_2\cos(4x_1) $ & $X_1 =\partial_3$ & $j=0$   & 4\cr
$F=e^{-x_2}(n_1\cos(4x_1) +n_2\sin(4x_1))$ & $X_2 =\partial_2+x_3\partial_3$ &
	$I_1= $& \cr
$G=e^{-2x_2}(n_0-n_1\sin(4x_1) +n_2\cos(4x_1)) $ &  $X_3 =e^{x_2}\partial_1+ 2x_3\partial_2  $  &$ -2n_0/\sqrt{n_1^2+n_2^2}   $ & \cr
$n_0,n_1,n_2 = const, ~ n_0^2 > n_1^2+n_2^2 \neq 0 $ &$~~~+(x_3^2-e^{-2x_2})\partial_3 $  &  & \cr
\mr     
\end{tabular}
\end{table}

\begin{table}[h]
\small        
\caption{\label{$G_3VIII.4$} Bianchi type $G_3$VIII.4} 
\lineup
\begin{tabular}{@{}*{4}{|l|l}}  
\br
metric\ & generators\ &  invariants\ & class \cr
\mr	
$E= 1$ & $X_1 =\partial_3$   &$j= \sqrt{1-1/k}$   & 2b \cr
$F= e^{-x_1}$ & $X_2 = \partial_1 +x_3\partial_3$ & 
$I_1 = -2/\sqrt{k},~I_2=I_3=J=0 $  & \cr
$G= k(x_2)e^{-2x_1}$ & $X_3=2x_3\partial_1+ x_3^2\partial_3$  &   
$L= k_{,2}e^{is/2}/\sqrt{2k(k-1)^2}$ &  \cr
  &   & $M=-ik_{,2}e^{is/2}/\sqrt{2k(k-1)^3}$  &  \cr
\mr     
\end{tabular}
\end{table}

\begin{table}[h]
\small        
\caption{\label{$G_3VIII.5$} Bianchi type $G_3$VIII.5 (intransitive group)} 
\lineup
\begin{tabular}{@{}*{4}{|l|l}}  
\br
metric\ & generators\ &  invariants\ & class \cr
\mr	
$E= a(x_1)$ & $X_1 =\partial_3$   &                    & 5 \cr  
$F= 0  $ & $X_2 = \partial_2 +x_3\partial_3$ &         & \cr 
$G= a(x_1)e^{-2x_2}$ & $X_3=2x_3\partial_2+ (x_3^2-e^{2x_2})\partial_3$ &  & \cr
\mr     
\end{tabular}
\end{table}

\begin{table}[h]
\small        
\caption{\label{$G_3IX.1$} Bianchi type $G_3$IX.1}
\lineup
\begin{tabular}{@{}*{4}{|l|l}}
\br
metric & generators &  invariants & class\cr
\mr	
$E= m(x_1) $ & $X_1 =\partial_3$   &         & 5 \cr
$F= 0 $  & $X_2 = \cos(x_3)\partial_2 + \sin(x_3)\tan(x_2)\partial_3 $ &  & \cr
$G= m(x_1)\cos^2(x_2) $& $X_3 = -\sin(x_3)\partial_2 +\cos(x_3)\tan(x_2)\partial_3 $  &  &\cr
\mr     
\end{tabular}
\end{table}

\begin{table}[h]
\small      
\caption{\label{$G_3IX.2$} Bianchi type $G_3$IX.2}
\lineup
\begin{tabular}{@{}*{4}{|l|l}}  
\br
metric & generators &  invariants & class\cr
\mr	
$E= l+m\sin(2x_1) +n\cos(2x_1) $ & $X_1 =\partial_3$   & $j= 0 $ & 4 \cr
$F= \cos(x_2)(n\sin(2x_1)- $  & $X_2 = \frac{\sin(x_3)}{\cos(x_2)}\partial_1 +              $ &  & \cr 
$~~~~~~m\cos(2x_1))  $ &$  \cos(x_3)\partial_2 +\sin(x_3)\tan(x_2)\partial_3$ &  $I_1=$ & \cr
$G= \cos^2(x_2)(l-m\sin(2x_1)-  $& 
$X_3 = \frac{\cos(x_3)}{\cos(x_2)}\partial_1 -$ & $\frac{2l}{\sqrt{m^2+n^2}} $  &\cr
$~~~~~~~n\cos(2x_1)) $& $  \sin(x_3)\partial_2 +\cos(x_3)\tan(x_2) \partial_3$     &   &  \cr
$l,m,n = const$ &   &   & \cr
\mr     
\end{tabular}
\end{table}

\begin{table}[h]
\small        
\caption{\label{$G_4V$}  $G_4$V}
\lineup
\begin{tabular}{@{}*{4}{|l|l}}  
\br
metric & generators & invariants & class\cr
\mr	
$E= ae^{-2x_1}$& $X_1 =\partial_3$     &      & 5\cr
$F= 0$ & $X_2 =\partial_2$   &    & \cr
$G= ae^{-2x_1}$& $X_3 =\partial_1 +x_2\partial_2+x_3\partial_3$  & & \cr
$a = const$ & $X_4 = x_3\partial_2-x_2\partial_3$     &  &  \cr
\mr     
\end{tabular}
\end{table}

\begin{table}[h]
\small        
\caption{\label{$G_4VII$} $G_4$VII}
\lineup
\begin{tabular}{@{}*{4}{|l|l}}  
\br
metric & generators & invariants & class\cr
&       &  &  \cr
\mr	
	$E= 1$ & $X_1 =\partial_3$ & $j= \sqrt{1-1/k}   $  & 2b\cr
$F= e^{-x_1} $  & $X_2 = \partial_1 + x_3 \partial_3 $ & $
I_1 = -\frac{2}{\sqrt{k}}, ~I_2=0,~I_3=0$ & \cr
$G= ke^{-2x_1}$ &  $X_3 = 2x_3\partial_1 + x_3^2\partial_3$  & $J=L=M=0 $& \cr
$k= const>1$ & $X_4 = \partial_2$ & &  \cr
\mr     
\end{tabular}
\end{table}

\clearpage



\section*{References}
\begin{thebibliography}{}

\bibitem{alvarez} 
F.~Fernandez-Alvarez, News tensor on null hypersurfaces, arXiv:2407.14909v1 [gr-qc] 20 Jul 2024

\bibitem{ashtekar}
A.~Ashtekar, S.~Speziale,
Null Infinity as a Weakly Isolated Horizon, arXiv:2402.177977v1 [hep-th] 28 Feb 2024

\bibitem{bartnik}
R.~Bartnik, Einstein equations in the null quasispherical gauge, Class. Quantum Grav. {\bf 14}, 2585-2194 (1997)

\bibitem{bekkara1}
E.~Bekkara, C.~Frances, and A.~Zeghib,
On lightlike geometry: isometric actions, and rigidity aspects,
C.R. Acad. Sci. Paris, Ser. I 343 (2006) 317-321

\bibitem{bekkara2}
E.~Bekkara, C.~Frances, and A.~Zeghib,
Actions of semisimple Lie groups preserving a degenerate riemannian metric,
Transactions AMS {\bf 362}, 2415-2435 (2010)

\bibitem{bianchi}
L.~Bianchi, Lezioni sulla teoria dei gruppi continui finiti di trasformazioni,
Spoerri, Pisa (1918)

\bibitem{blitz} 
S.~Blitz, D.~McNutt, 
Horizons that Gyre and Gimble: A Differential Characterization of Null Hyersurfaces, arXiv:2310.08141v2 [gr-qc] 18 Mar 2024

\bibitem{bondi}
H.~Bondi, M.~van den Berg, and A.~Metzner,
1962 Proc.R.Soc. {\bf A 269}, 21

\bibitem{chandra} 
V.~Chandrasekaran, E.~Flanagan and K.~Prabhu,
Symmetries and charges of general relativity at null boundaries, J. High Energy Physics (2018) 1-64, arXiv:1807.11499v2 [hep-th] 12 Nov 2018

\bibitem{daut1}
G.~Dautcourt, Zum charakteristischen Anfangswertproblem der Einsteinschen Feldgleichungen, Annalen der Physik, {\bf 467}, 302 (1963).

\bibitem{daut2} 
G.~Dautcourt, Isotrope Flaechen in der allgemeinen Relativitaetstheorie, Habilschrift, Humboldt-Universitaet Berlin, Dezember 1965

\bibitem{daut3} 
G.~Dautcourt, Characteristic Hypersurfaces in General Relativity, J. Mathematical Physics {\bf 7}, 1492-1501 (1966).

\bibitem{daut4} 
G.~Dautcourt, Isotropic hypersurfaces in General Relativity admitting groups of motions, Acta Physica Polonica, {\bf B 11},791 (1980)

\bibitem{eisenhart1}
L.P.~Eisenhart, (1927) Non-Riemannian geometry (Amer. Math. Soc., ProvIdence, Rhode Island)

\bibitem{eisenhart2}
L.P.~Eisenhart, (1933) Continous Groups of Transformations (Princeton University Press, Princeton 1933)

\bibitem{fubini}
G.~Fubini, Mem. di Mat. (3) {\bf 8}, 39 (1903)

\bibitem{katsuno}
K.~Katsuno, Null hypersurfaces in Lorentzian Manifolds II,
Math. Proc. Cambridge Phil. Soc. {\bf 89}, 525-532 (1981)

\bibitem{kamke}  
E.~Kamke, Differentialgleichungen L\"osungsmethoden und L\"osungens 
II. Partielle Dfferentialgleichungen erster Ordnung f\"ur eine gesuchte Funktion (Geest \& Portig K.G., Leipzig 1956)


\bibitem{lemmer}
G.~Lemmer, On Covariant Differentiation Within a Null Hypersurface, 	
Il Nuovo Cimento {\bf 37}, 1659-1672 (1965)

\bibitem{manz1}
M.~Manzano, M.~Mars, The constraint tensor for null hypersurfaces,
arXiv:2309.14813v2 [gr-qc] 31 Jul 2024

\bibitem{manz2}
M.~Manzano, M.~Mars, Embedded Hypersurface Data and Ambient Vector Fields: 
Abstract Killing Horizons of order zero/one, arXiv:2403.12228v1 [gr-qc] 18 Mar 2024

\bibitem{mars1} 
M.~Mars, Abstract null geometry, energy-momentum map and applications to the constraint tensor, arXiv:2402.07488v1 [gr-qc] 12 Feb 2024

\bibitem{My} 
S.B.~Myers, Isometries of 2-dimensional Riemannian manifolds into themselves,
Proc. Natl. Acad. Sci. U.S.A. {\bf 22}, 297-300 (1936)

\bibitem{newman}
E.~Newman, R.~Penrose, An approach to gravitational radiation by a method of spin coefficients,	J. Mathematical Physics {\bf 3}, 566-578 (1962).

\bibitem{odak}
G.~Odak, A.~Rignon-Bret, S.~Speziale, 
General gravitational charges on null hypersurfaces, arXiv:2309.03854v1 [gr-qc] 7 Sep 2023.

\bibitem{palomo}
F.~Palomo, Lightlike manifolds and Cartan Geometries, 
 arXiv:2003.09448v1 [math.DG] 20 Mar 2020

\bibitem{pen0}
R.~Penrose. Null hypersurface initial data for classical fields of arbitrary spin and general relativity. Aerospace Research Laboratories Report 63-56 ed. P. G. Bergmann. reprinted Gen. Rel. Grav. 12:225, 198.

\bibitem{pen1}
R.~Penrose, Les Houches Lectures 1963, in: \emph{Relativity, Groups and Topology}, ed. C.M.~DeWitt \& B.S.~DeWitt, Blackie \& Son, London 1964.



\bibitem{pen2}
R.~Penrose, The geometry of impulsive gravitational waves,
		in: \emph{General Relativity},ed. L. O'Raifeartaigh, 
		Clarendon, Oxford, pp. 101-115~ (1973).

\bibitem{petrov}
A.Z.~Petrov, Einstein Spaces, Pergamon Press 1969.

\bibitem{sachs}
R.K.~Sachs, On the Characteristic Initial Value Problem in Gravitational Theory, J. Mathematical Physics {\bf 3}, 908-914 (1962).



\bibitem{smarr}
L.~Smarr, Surface Geometry of Charged Black Holes, Phys. Rev. {\bf D 7}, 289-295 (1973).

\bibitem{MMM}
H.~Stephani, D.~Kramer, M.~MacCallum, C.~Hoenselaers and E.~Herlt, 
\emph{Exact Solutions of Einstein's Field Equations} (Cambridge University Press, Cambridge, UK,  2003), 2nd ed.

\bibitem{wald}
R.M.~Wald, \emph{General Relativity}  (The University of Chicago Press, Chicago 1984).


\end{thebibliography}
\end{document}